\documentclass{gPAA2e}
\usepackage{rotating}
\usepackage[algo2e, ruled, vlined, commentsnumbered, linesnumbered]{algorithm2e}
\usepackage{amsmath}
\usepackage{epstopdf}
\usepackage{color}
\usepackage{url}

\begin{document}

\title{An Efficient Multi-core Implementation of the Jaya Optimisation Algorithm}

\author{Panagiotis D. Michailidis \\
Department of Balkan, Slavic and Oriental Studies, University of Macedonia\\
156 Egnatia str., 54636 Thessaloniki, Greece\\
E-mail: pmichailidis@uom.gr}

\maketitle

\begin{abstract}
In this work, we propose a hybrid parallel Jaya optimisation algorithm for a multi-core environment with the aim of solving large-scale global optimisation problems. The proposed algorithm is called HHCPJaya, and combines the hyper-population approach with the hierarchical cooperation search mechanism. The HHCPJaya algorithm divides the population into many small subpopulations, each of which focuses on a distinct block of the original population dimensions. In the hyper-population approach, we increase the small subpopulations by assigning more than one subpopulation to each core, and each subpopulation evolves independently to enhance the explorative and exploitative nature of the population. We combine this hyper-population approach with the two-level hierarchical cooperative search scheme to find global solutions from all subpopulations. Furthermore, we incorporate an additional updating phase on the respective subpopulations based on global solutions, with the aim of further improving the convergence rate and the quality of solutions. Several experiments applying the proposed parallel algorithm in different settings prove that it demonstrates sufficient promise in terms of the quality of solutions and the convergence rate. Furthermore, a relatively small computational effort is required to solve complex and large-scale optimisation problems. 
\end{abstract}


\section{Introduction}
In order to solve complex and difficult problems in science, engineering, economics, and other fields, researchers create models and reduce problems to well-studied optimisation problems \cite{Talbi:2009:MDI:1718024}. An optimisation problem can be defined as follows. Given an objective function, a set of decision variables, and a set of constraints that limit the values of the variables, the aim is to determine feasible solutions of the variables that minimise or maximise the value of the objective function. 

In many scientific applications, optimisation problems lead to complex and non-linear objective functions with large numbers of variables \cite{Rao2011303}. In these complex and large-scale optimisation problems, the search space is large and multi-dimensional. Such problems cannot be solved using traditional numerical methods, such as linear programming and dynamic programming, because these methods cannot be employed to find the global optimal solution. Therefore, there is a need for efficient and effective optimisation methods. Several modern population-based metaheuristic optimisation methods have been developed over the past two decades. These methods can mainly be classified into two categories: evolutionary algorithms (EA) and swarm intelligence (SI) based algorithms. Examples of popular evolutionary algorithms are the genetic algorithm (GA) \cite{Goldberg:1989:GAS:534133}, differential evolution (DE) \cite{Mezura-Montes:2010:DEC:1857270.1857549}, and the artificial immune algorithm (AIA) \cite{Farmer:1986:ISA:25201.25213}. Some well-known swarm intelligence based algorithms are particle swarm optimisation (PSO) \cite{kennedy:1995:PartSwarmOpt}, ant colony optimisation (ACO)  \cite{Dorigo:2004:ACO:975277}, and artificial bee colony optimisation (ABC) \cite{Karaboga20113021}. Detailed information regarding the above-mentioned population-based metaheuristics can be found in \cite{Boussaïd201382}. The main drawback of the above algorithms is that they employ their own algorithm-specific parameters (such as mutation and crossover rate for GA or inertia weight and social cognitive parameters for PSO), which require extensive tuning before performing computational experiments in order to achieve a superior performance.

Recently, Rao \cite{Rao2011303, jaya12016} introduced two new population-based metaheuristics, known as teaching-learning based optimisation (TLBO) and Jaya, which only require common control parameters (such as population size and a maximum number of iterations or generations), and not any algorithm-specific parameters. The TLBO algorithm \cite{Rao2011303,Rao20121,Rao2013710,Satapathy201428} imitates the two phases of a traditional classroom scenario, teaching and learning. A detailed explanation of the TLBO algorithm can be found in \cite{Rao2011303}. Several TLBO variants and applications are also available in the research literature \cite{Rao:2015:TLB:2857289, tlborao2016}. Unlike the two phases of the TLBO algorithm, the Jaya algorithm \cite{jaya12016,doi:10.1080/0305215X.2016.1164855} uses only the teacher phase, and is comparatively simple to apply. It is based on the fact that a solution can be obtained for a given problem that moves towards the best solution and avoids the worst solution. In this work, we put a particular emphasis on the Jaya algorithm, because it constitutes a straightforward and powerful global optimisation algorithm with few parameters, is easy to implement, and has been applied to solve various optimisation problems effectively and efficiently \cite{doi:10.1080/0305215X.2016.1164855,Rao2016572,Rao2016,Rao2016b,en9090678}. One of the main advantages of using the Jaya algorithm is that it prevents solutions from becoming trapped in local optima, unlike many other population-based metaheuristics. This notable feature of the Jaya method makes it superior to other population-based optimisation methods \cite{en9090678}.

The main issues regarding metaheuristics, including Jaya algorithm, is that it increases the amount of time as the size of the search space increases and it also suffers from premature convergence toward a local optimum when solving complex and high dimensional optimisation problems. Parallel computing techniques can be employed not only to improve the search time but also to improve the quality of solutions of the optimisation problems \cite{ITOR:ITOR862}. In the research literature, four popular parallel models have been identified to improve the computational efficiency of population-based metaheuristic algorithms. These models are parallel independent runs, master-worker, multi-population cooperative master-worker, and cellular models \cite{Pedemonte20115181, Luque:2013:PGA:2564896}. The parallel independent runs, master-worker, and cellular models achieve high computational speedups when implementing parallel metaheuristics, but they produce a similar quality of solutions compared with sequential metaheuristic algorithms. On the other hand, the multi-population cooperative master-worker models (coarse grained) produce a better quality of solutions with less computational effort. The multi-population cooperative models consisted of multiple populations where each population in one processor of the parallel platform evolves according to a metaheuristic algorithm independently and exchanges the solutions of the algorithm with other populations by sporadic migrations. 

Some recent research efforts on parallelisation of population-based metaheuristics have been developed on parallel systems in the context of the multi-population cooperative paradigm. More specifically, researchers in \cite{Parpinelli2011} proposed three parallel models for the ABC algorithm,  master-worker, multi-hive with migrations and hybrid hierarchical and they also conducted an peformance evaluation on a cluster of workstations using MPI programming paradigm. In \cite{subotic2014}, they parallelised ABC algorithm on a multi-core platform with a multiple swarm model where whole colony is divided into four sub-colonies. There was a periodic communication of the solutions among the sub-colonies. In \cite{Umbarkar2014936}, they implemented a dual population genetic algorithm on multi-core platform in order to solve the problem of premature convergence. In \cite{Basturk:2013:PAC:2535050.2535175}, they implemented a multi-population model based parallel ABC algorithm in MPI and investigated an extended performance study about different migration parameters and communication topologies among the subpopulations. Recently, a hierarchical cooperative search scheme has been reported inspired by the natural ecosystem \cite{Parpinelli2015}. The ecosystem as a whole can be composed by populations and each population running a metaheuristic algorithm. These populations are scattered dynamically in ecological habitats according to a hierarchical clustering algorithm.  By this way, there are two levels of communications, intra-habitats and inter-habitats communication, that favors co-evolution. These two communications can be implemented through ecological symbiotic relationships \cite{Parpinelli2015}. The advantage of this ecosystem is that there is a better balance between exploration and exploitation ability and making the algorithm perform well in complex optimisation problems. 

The aforementioned cooperative search models, the number of populations is limited by the number of processing elements in the parallel platform or the number of the populations is small and, the cooperative between populations is based on the different network and migration topologies or symbiotic relationships. Most of these studies introduced parallelisation schemes with a relatively small number of processors. 
Another issue regarding the above-mentioned parallel research studies is that they have considered small and medium sized dimensional problems in their computational experiments.
Furthermore, most of these studies have been concentrated on the parallelisation of the search space on the basis of only the population size. Therefore, these studies suffer from the curse of dimensionality problem when applied to high dimensional problems. In other words, the search time increases significantly as the dimension of search space increases. Finally, applying the multi-population model to the Jaya algorithm is possible to arise the problem of the slow convergence speed in solving complex and high dimensional problems. The convergence speed of an optimisation algorithm is a very important characteristic to reach the global optimal solution with a very high rate.

Motivated by the performance advantages of the multi-population model and the limitations of the aforementioned research studies, in this paper, we propose an enhanced software-based hierarchical hyper-population cooperative parallel Jaya optimisation algorithm (HHCPJaya) on a multi-core environment. The main benefits of using the proposed parallel implementation is the possibility of further improving the quality of solutions, the convergence speed and to avoid premature convergence with much faster execution times for solving large-scale optimisation problems. The main contribution of this work is that we incorporate three characteristics into the proposed parallel master-worker algorithm. First, we utilise hyper-population approach instead of multi-population, where the original population is divided into many small and independent local subpopulations over the population size and the dimensions of the problem, simultaneously. Each of which focuses on a distinct subset of the original population dimensions. In this case, the number of subpopulations is significantly larger than the number of cores of the multicore platform, with multiple independent subpopulations assigned to each core. Each subpopulation evolves independently for one generation of the sequential Jaya algorithm. This can enhance the exploration capability of the search space, and further improve the quality of solutions. Second, we incorporate a simplified two-level local and global hierarchical cooperative search of solutions among the subpopulations introduced at each iteration of the parallel algorithm, instead of the periodical cooperation that was applied in the original multi-population approaches. In this two-level search scheme, all workers determine local solutions from their respective local subpopulations at the bottom level, and these solutions interact with the master to determine the global solutions at the top level. Third, an updating phase is performed on solutions on the respective local subpopulations, based on global solutions, in order to move the solutions of all local subpopulations further towards the global solution. Note that this  phase is new except for the process of updating based on local solutions, which is in the original Jaya algorithm. These characteristics enhance the explorative and exploitative nature of the hyper-population, and thus help to significantly improve the quality of solutions, convergence speed, and execution time of the parallel algorithm for solving high dimensional problems.  

To the best of our knowledge, the proposed multi-core implementation of the Jaya algorithm with the above-incorporated characteristics has not been reported in the existing literature. Finally, the proposed parallel algorithm was implemented on a multi-core platform using the OpenMP programming interface, and was evaluated for some standard benchmark function problems by studying the effects of parameters such as the dimension size, the number and size of subpopulations, and problems with different complexities. The results demonstrate that the proposed parallel approach is efficient and fast in terms of the convergence rate, quality of solutions, and execution times for solving large-scale optimisation problems. More specifically, the number of small subpopulations used in the hyper-population approach has a significant effect on the performance of the parallel algorithm.

The remainder of this paper is organised as follows. Section~\ref{seqjaya} provides a description of the original sequential Jaya algorithm. The proposed multi-core implementation of the Jaya algorithm based on the hierarchical hyper-population cooperative master-worker model is explained in detail in Section~\ref{parjaya}. Experimental results are presented in Section~\ref{results}. Finally, some conclusions and future research plans are discussed in Section~\ref{conclusions}.

\section{Sequential Jaya Algorithm}
\label{seqjaya}
The Jaya algorithm is a population-based metaheuristic for solving optimisation problems, originally proposed by Rao \cite{jaya12016,doi:10.1080/0305215X.2016.1164855,jayaonline}. The basic idea of the algorithm is that it always tries to improve the solution by avoiding the worst solution at every iteration. By using this mechanism, the algorithm aims to be victorious, hence the name Jaya, which is derived from a Sanskrit word meaning `victory' \cite{jaya12016}. Algorithm~\ref{alg-jaya} presents the framework of the Jaya algorithm.

\begin{algorithm2e}
\DontPrintSemicolon
\KwIn{Population size $n$, number of variables $m$, limits of variables ($X_{min}, X_{max}$), maximum number of iterations $max\_iter$}
\KwOut{Best solution}
	$iteration \leftarrow 0$\;
	Initialize\_population($X$, $n$, $m$)\;
	Evaluate\_population($X$, $n$, $m$, $fv$)\;
	\Repeat{$iteration \neq max\_iter$}{
		Memorise\_best\_and\_worst\_solution\_in\_the population($X$, $n$, $m$, $fv$, $best$, $worst$)\;
		Update\_the\_population\_of\_solutions($X$, $n$, $m$, $fv$, $best$, $worst$)\;
		$iteration \leftarrow iteration + 1$\;
	}
	Print\_the\_best\_solution\;
	\caption{Sequential Jaya algorithm}
	\label{alg-jaya}
\end{algorithm2e}

First, the algorithm accepts the control input parameters, such as the population size $n$, number of variables $m$, lower and upper limits of variables ($X_{min}, X_{max}$), and maximum number of iterations $max\_iter$. The algorithm also accepts the objective function $f(x)$ of the optimisation problem. Then, the algorithm begins with the initialisation of the population, in the search space that represents solutions (line 2). The population of the solutions is represented by an $n \times m$ matrix $X_{i,j}$, where $n$ is the population size, or number of candidate solutions, and $m$ is the number of variables or dimensions. The population matrix for the algorithm is expressed as follows:

$X_{i,j}=\begin{bmatrix}
X_{11} & X_{12} &  \dots  & X_{1m} \\
X_{21} & X_{22} &  \dots  & X_{2m} \\
\vdots & \vdots & \ddots & \vdots \\
X_{n1} & X_{n2} &  \dots  & X_{nm}
\end{bmatrix}$

\noindent where $i=1,2,\dots,n$ and $j=1,2,\dots,m$. The values of the matrix $X$ should be within the limits of the parameters, $X_{min} \le X_{i,j} \le X_{max}$. During the initialisation phase, the population matrix is randomly generated using equation~\ref{eq1}, within the limits of variables to be optimised:
\begin{equation}
X_{i,j}= X_{min}+rand(0,1)(X_{max}-X_{min})
\label{eq1}
\end{equation}

\noindent where $i=1,2,\dots,n$ and $j=1,2,\dots,m$. Here, rand(0,1) is a random number in the range [0,1]. 
% can be obtained from a normal distribution function. 

The next step is to evaluate the population of solutions (line 3 of Algorithm~\ref{alg-jaya}). In this step, the algorithm calculates the fitness value of each candidate solution of the population based on the objective function $f$. The fitness values for the $n$ candidate solutions of the population are stored in a vector $fv$ of size $n$.

At each iteration, the Jaya algorithm performs two steps: memorising the best and worst solutions in the population (line 5 of Algorithm~\ref{alg-jaya}) and updating the population of solutions based on the best and worst solutions (line 6 of Algorithm~\ref{alg-jaya}). During the memorisation step, the algorithm examines the values of the fitness vector $fv$ in the entire population and selects the best and the worst fitness values. The best and the worst candidate solutions that correspond to these values are stored in the vectors $best$ and $worst$ of size $m$, respectively. The working procedure of the memorisation step is shown in Algorithm~\ref{algo4}.

\begin{algorithm2e}
	\DontPrintSemicolon
	\KwIn{$X$: population matrix, $n$: population size, $m$: the number of variables, $fv$: fitness vector}
	\KwOut{$best, worst$: the best and worst solution vectors}
	$bestf \leftarrow LONG\_MAX$; $worstf \leftarrow LONG\_MIN$\;
	\For{$i\leftarrow 0$ \KwTo $n$}{
		  \If{$fv_{i} < bestf$} {
		  	$bestf \leftarrow fv_{i}$\;
		  	\For{$j\leftarrow 0$ \KwTo $m$}{
			  		$best_{j} \leftarrow X_{i,j}$
			}
		  }
		   \If{$fv_{i} > worstf$} {
		   	$worstf \leftarrow fv_{i}$\;
		   	\For{$j\leftarrow 0$ \KwTo $m$}{
		   		$worst_{j} \leftarrow X_{i,j}$
		   	}
		   }
	}
	\caption{Memorise\_best\_and\_worst\_solution\_in\_the population}
	\label{algo4}
\end{algorithm2e}

In the updating phase, a new solution is produced for each candidate solution, as defined in equation~\ref{eq2}:

\begin{equation}
X^{new}_{i,j}=X_{i,j} + r_{1}(best_{j}-|X_{i,j}|) - r_{2}(worst_{j}-|X_{i,j}|)
\label{eq2}
\end{equation}

\noindent where $i=1,2,\dots,n$, $j=1,2,\dots,m$, $X^{new}_{i,j}$ is the updated value of $X_{i,j}$, and $r_{1}$ and $r_{2}$ are the two random numbers in the range [0,1]. The term $r_{1}(best_{j}-|X_{i,j}|)$ indicates the tendency of the solution to move closer to the best solution, and the term $- r_{2}(worst_{j}-|X_{i,j}|)$ indicates the tendency of the solution to avoid the worst solution \cite{jaya12016,doi:10.1080/0305215X.2016.1164855}. At the end of the updating phase, the algorithm examines whether the solution corresponding to $X^{new}_{i,j}$ gives a better fitness function value than that corresponding to $X_{i,j}$, then either accepts and replaces the previous solution or keeps the previous solution, depending on the outcome. All of the accepted solutions at the end of the current iteration are maintained, and these solutions become the input to the next iteration. The working procedure of the updating phase is shown in Algorithm~\ref{algo5}.

\begin{algorithm2e}
	\DontPrintSemicolon
	\KwIn{$X$: population matrix, $n$: population size, $m$: the number of variables, $fv$: fitness vector, $best, worst$: the best and worst solution vectors}
	\KwOut{$X$: updated population matrix, $fv$: updated fitness vector}
	\For{$i\leftarrow 0$ \KwTo $n$}{
		\For{$j\leftarrow 0$ \KwTo $m$}{
				$X^{new}_{j}=X_{i,j} + r_{1}(best_{j}-|X_{i,j}|) - r_{2}(worst_{j}-|X_{i,j}|)$
			}
		$fnew \leftarrow f(X^{new})$\;
		  \If{$fnew < fv_{i}$} {
		  	\For{$j\leftarrow 0$ \KwTo $m$}{
		  		$X_{i,j} \leftarrow X^{new}_{j}$
		  	}
		  	$fv_{i} \leftarrow fnew$\;
		  }

	}
	\caption{Update\_the\_population\_of\_solutions}
	\label{algo5}
\end{algorithm2e}

The two algorithms mentioned above (algorithms~\ref{algo4} and~\ref{algo5}) will continue to execute until the number of iterations or generations reaches its defined maximum value.

\section{Hierarchical Hyper-population Cooperative Parallel Jaya Algorithm}
\label{parjaya}
In this section, a hybrid parallel Jaya algorithm called HHCPJaya is proposed for improving the quality of solutions and the convergence rate within a reasonable amount of time. This algorithm unifies the hyper-population approach with a two-level hierarchical cooperation model among the subpopulations. In the following, a detailed description of HHCPJava will be provided. Finally, this parallel algorithm is implemented using the master-worker programming paradigm as described later in this section.

\subsection{High-Level Description of HHCPJaya}
\begin{figure}
	\includegraphics[width=\textwidth]{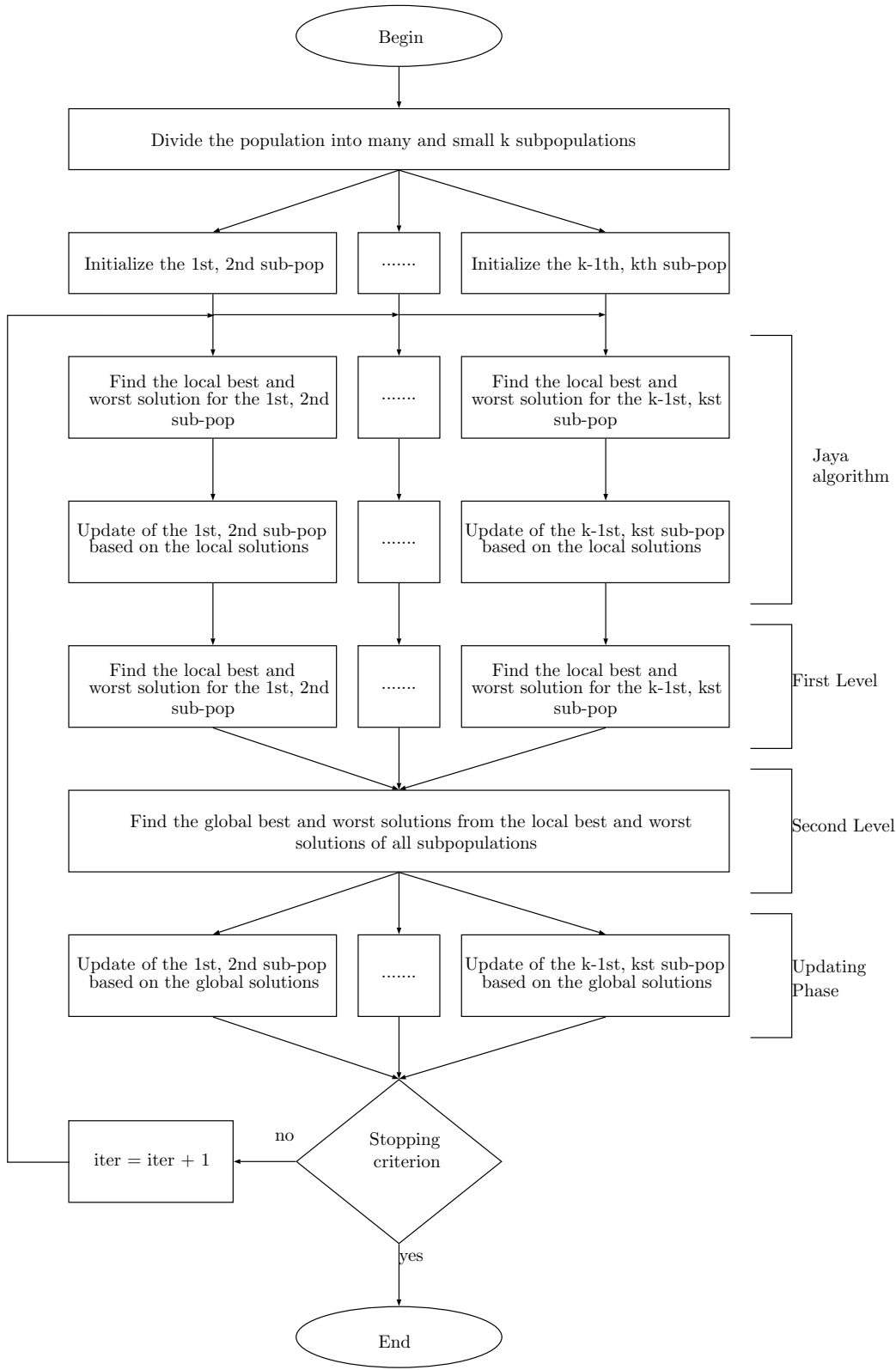}
	\caption{The overall structure of the HHCPJaya algorithm}
	\label{algorithm-flow}
\end{figure}

The main underlying idea of the proposed algorithm is that the original population is divided into many local and independent subpopulations, using a general block decomposition approach. Each subpopulation focuses on a distinct subset of the initial population dimensions. In this case, the number of subpopulations is significantly greater than the number of worker threads, with multiple independent subpopulations assigned to each different worker thread. This approach is called hyper-population, and it is used to increase the diversity of the population and avoid trapping solutions in local optima. Therefore, the proposed hyper-population approach improves the exploration capability. This decomposition procedure for the hyper-population is explained in further detail later in this section. During each iteration of the HHCPJaya algorithm, each worker thread performs the memorisation and updating steps of the Jaya algorithm for one generation on each of its local subpopulations, independently and sequentially, in its local memory. In other words, each thread finds the best and worst local solutions for each local subpopulation, and updates the respective subpopulation based on these local solutions. Then, we introduce a two-level hierarchical cooperative search of solutions among subpopulations, in order to further improve the quality of solutions. More specifically, each worker determines the best and worst local solutions from all of its local subpopulations at the first level of the hierarchy. These local solutions from each worker interact with the master thread at the second level of the hierarchy to determine the best and worst global solutions. At the end of each iteration, these global solutions are made available to all of the worker threads, and then each thread performs the updating phase on its respective local subpopulations based on the global best and worst solutions. We also introduce this additional updating phase on the respective local subpopulations, to improve further the solutions on each of the local subpopulations towards the best solutions, thus further avoiding the worst solutions. This enhances the exploitation capability of the subpopulations and the convergence speed of the HHCJaya algorithm. We iterate the above procedure until the number of iterations reaches its defined maximum value. The hyper-population approach and the two-level hierarchical cooperative search scheme can improve the exploration and exploitation process on the search space, and may lead to reaching an optimal solution. The overall structure of the HHCPJaya algorithm is shown in Figure~\ref{algorithm-flow}. For simplicity, we assume that each worker thread processes two subpopulations.

\subsection{The Block Decomposition Approach}
\begin{figure}
	\includegraphics[width=\textwidth]{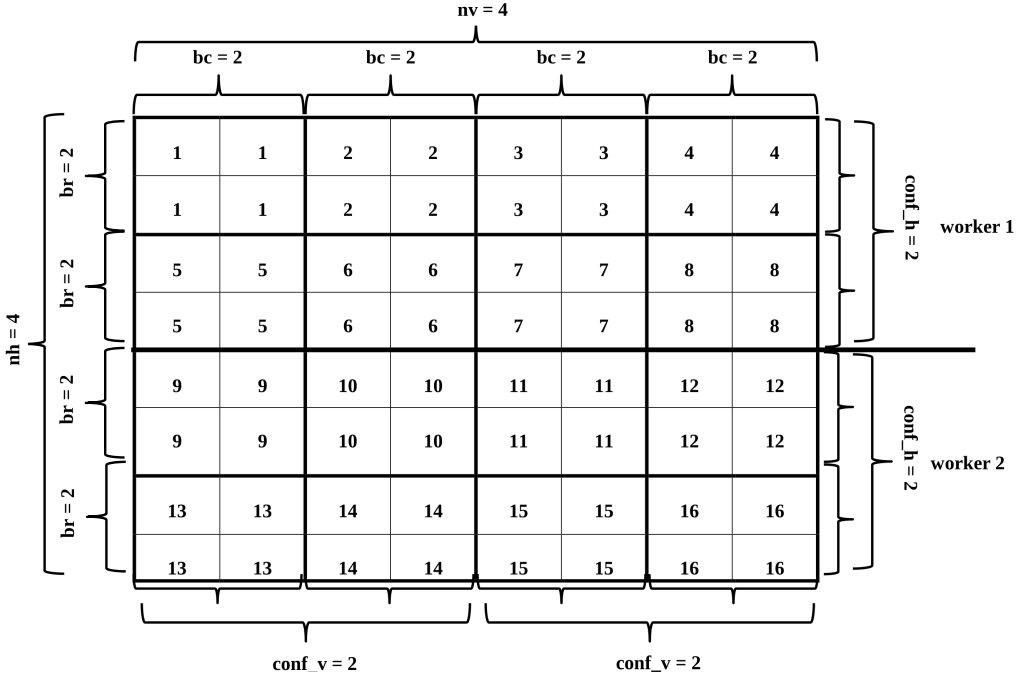}
	\caption{Conceptual illustration of the block partitioning strategy of a population matrix $X$ of size $8 \times 8$ among two worker threads}
	\label{data1}
\end{figure}

\begin{figure}
	\includegraphics[width=\textwidth]{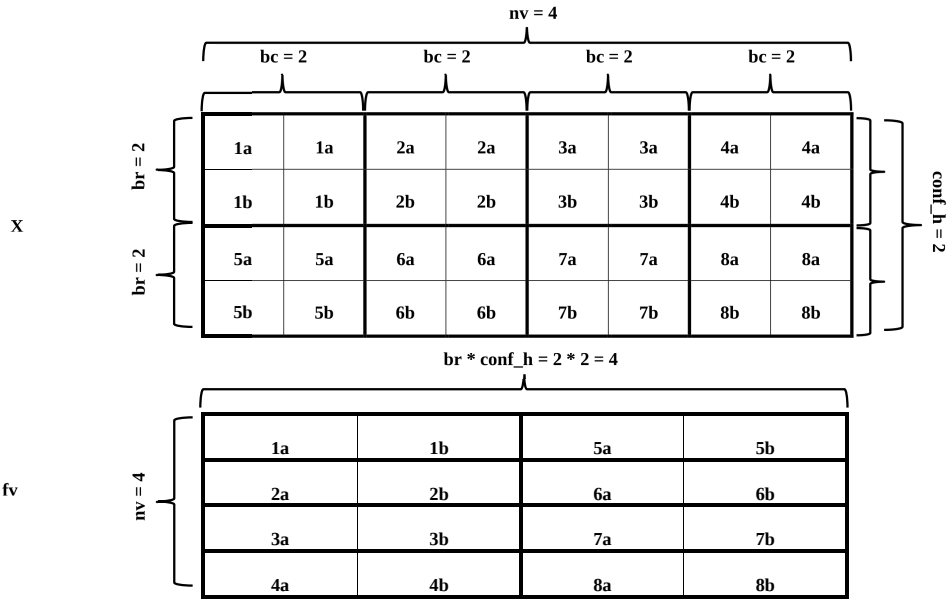}
	\caption{Conceptual illustration of the sizes of the matrices $X$ and $fv$ required by each worker thread to process the $l=8$ subpopulations}
	\label{data2}
\end{figure}

The general block decomposition strategy in the hyper-population approach is as follows. The original population matrix $X$ of size $n \times m$ is divided into a number of horizontal subpopulations $nh$ and a number of vertical subpopulations $nv$. The number of horizontal and vertical subpopulations are controlled by two multiplier coefficients $conf\_h$ and $conf\_v$, which are natural numbers expressing the number of divisions to be formed for the rows and columns of $X$, respectively, over the number of worker threads. Therefore, the number of horizontal and vertical subpopulations can be defined by the following relations: 
\begin{equation}
nh=conf\_h \cdot threads
\label{eqnh}
\end{equation}
\begin{equation}
nv=conf\_v \cdot threads
\label{eqnv}
\end{equation}
When $conf\_v \ge 1$ and $conf\_h \ge 1$, the hyper-population approach increases the numbers of horizontal and vertical subpopulations to be greater than the number of worker threads. The total number of resulting subpopulations $k$ is defined by the relation
\begin{equation} 
k= nv \cdot nh
\label{eqk}
\end{equation} 
and the size of all subpopulations is equal to $br\times bc$, where 
\begin{equation}
br=\frac{n}{nh}
\label{eqbr}
\end{equation} and 
\begin{equation}
bc=\frac{m}{nv}
\label{eqbc}
\end{equation}
In the HHCPJaya approach, the $k$ subpopulations are distributed among the worker threads by a classical blocks mapping, i.e., $\frac{nh}{threads} \cdot nv$. Therefore, a number $conf\_h$ of $nv$ vertical subpopulations are statically assigned to each worker thread, because the workload associated with each subpopulation is the same. Therefore, each thread has $l=conf\_h \times nv$ local subpopulations and processes a subpopulation of size $br \times bc$ each time. Figure~\ref{data1} illustrates the block decomposition of a population matrix of size $8\times 8$ into subpopulations when the values of parameters $conf\_h$ and $conf\_v$ are equal to 2 and the number of threads is two. Using the equations~\ref{eqnh},~\ref{eqnv},~\ref{eqk},~\ref{eqbr} and~\ref{eqbc}, we obtain $k=16$ subpopulations of size $2 \times 2$ and these are numbered as shown in Figure~\ref{data1}. In the same figure, we illustrate the static distribution of 16 subpopulations between two worker threads. More specifically, $l=8$ subpopulations are allocated to each worker thread.
In this case, each worker thread requires four new data structures to perform the Jaya algorithm for one iteration on each of its $l$ independent and local subpopulations. More specifically, each thread requires a new matrix $X$ of total size $(conf\_h \cdot br) \times m$ and a new fitness matrix $fv$ of total size $nv \times (conf\_h \cdot br)$ to store the solutions and the fitness vectors of $l$ independent and local subpopulations, respectively. Figure~\ref{data2} shows the sizes of the matrices $X$ and $fv$ required by each worker thread.
In the top part of the figure, we show the matrix $X$ that stores the $br$ solutions of size $bc$ for eight subpopulations, whereas the bottom part shows that each row of the matrix $fv$ stores two fitness vectors of size $br$ for the first two vertical subpopulations of the matrix $X$, as indicated by the parameter $conf\_h$. For instance, the elements $1a$ and $1b$ of the matrix $fv$ from Figure~\ref{data2} store the fitness values, and these are computed based on the $bc$ elements of the first and second rows, respectively, of the matrix $X$, which corresponds to the first subpopulation. Similarly, the other elements of the matrix $fv$ are computed based on the respective $bc$ elements of other subpopulations of the matrix $X$, as shown in Figure~\ref{data2}.
\begin{figure}[b]
	\includegraphics[width=\textwidth]{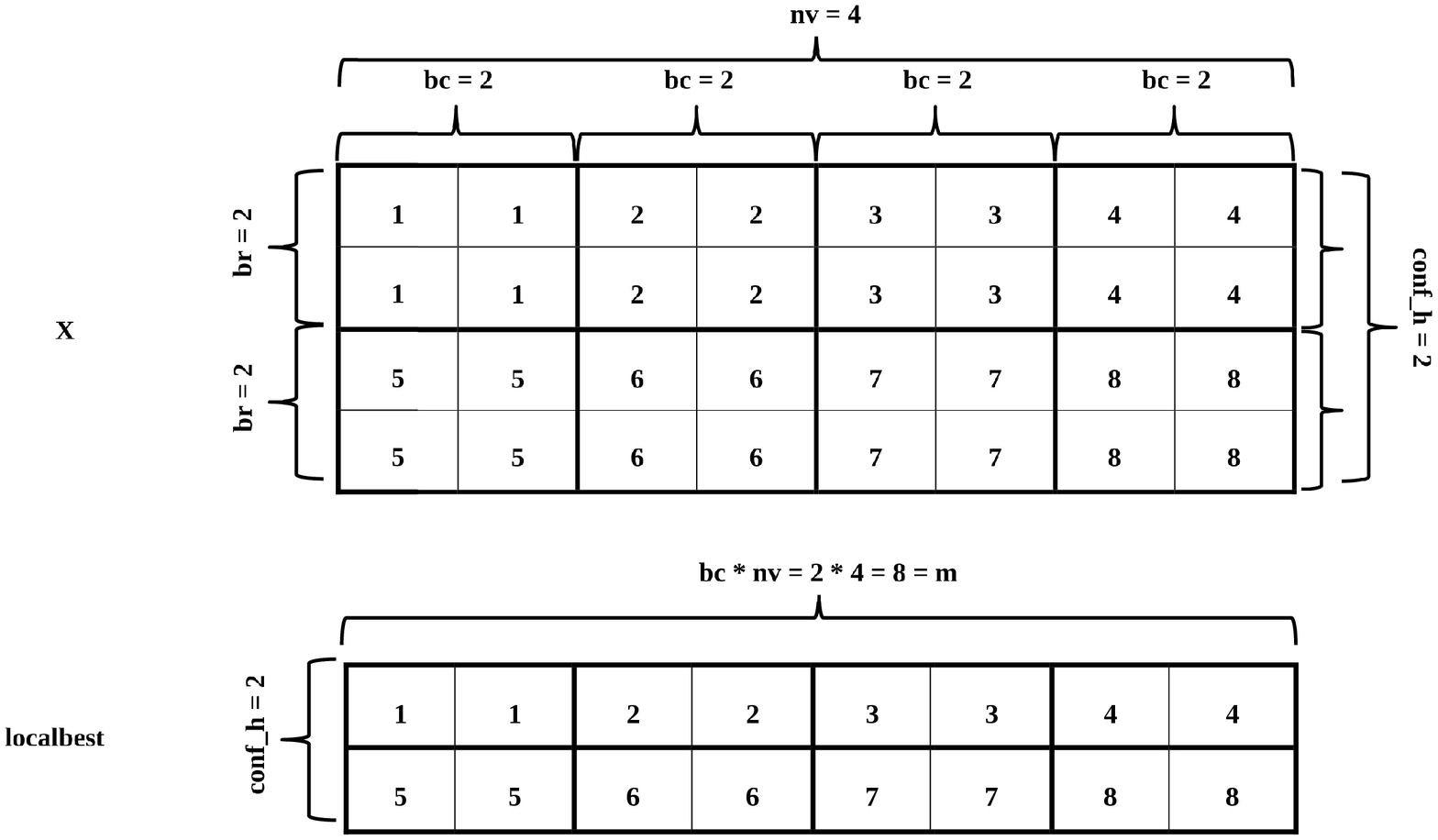}
	\caption{Conceptual illustration of the sizes of the matrices $X$ and $localbest$ required by each worker thread}
	\label{data3}
\end{figure}
Furthermore, each thread requires two new matrices $localbest$ and $localworst$ of total size $conf\_h \times m$ to store the local best and worst solutions of size $bc$ for each of the $l$ local subpopulations. Figure~\ref{data3} shows the sizes of the matrix $localbest$ required by each worker thread. 
In the top part of the figure, we depict the population matrix $X$ for each thread for illustrative purposes, and in the bottom part, we show the matrix $localbest$ that stores the best solutions of size $bc=2$ for the eight subpopulations of the matrix $X$. For instance, the first two elements of the matrix $localbest$ store the best solution of size $bc=2$ corresponding to the first subpopulation of the matrix $X$, as numbered and shown in Figure~\ref{data3}. Moreover, the size and data structure of the matrix $localworst$ are similar to those of $localbest$.
Note that the computations of the Jaya algorithm for each independent local subpopulation are performed on the different respective subsets of the original data structures $X$, $fv$, $localbest$, and $localworst$.

\subsection{Master - Worker Implementation Details}
The idea of combining the hyper-population approach using the block decomposition technique and the two-level hierarchical cooperative search of solutions among subpopulations leads to a coarse-grain master-worker model.
\begin{algorithm2e}
	\DontPrintSemicolon
	\KwIn{Population size $n$, number of variables $m$, horizontal partition parameter $conf\_h$, vertical partition parameter $conf\_v$, limits of variables ($X_{min}, X_{max}$), maximum number of iterations $max\_iter$, number of threads $threads$}
	\KwOut{Best solution}
	$iteration \leftarrow 0$; $nh \leftarrow conf\_h * threads$; $nv \leftarrow conf\_v * threads$; \\
	\ForEach{thread tid, in parallel}{
		Initialize\_population($X$, $nv$, $nh$)\;
		Evaluate\_population($X$, $fv$,  $nv$, $nh$)\;
		\Repeat{$iteration \neq max\_iter$}{
			Local\_memorize\_best\_worst\_sol($fv$, $X$, $localbest$, $localworst$,  $nv$, $nh$)\;
			barrier()\;
			Parallel\_update\_the\_population\_of\_sols($X$, $n$, $m$, $fv$, $localbest$, $localworst$, $nh$, $nv$)\;
			Local\_memorize\_best\_worst\_sol($fv$, $X$, $localbest$, $localworst$,  $nv$, $nh$)\;
			Local\_cooperative\_memorize\_best\_worst\_sol($fv$, $localbest$, $localworst$, $localbest\_thread$, $localworst\_thread$,  $nv$, $nh$)\;
			barrier()\;
			\If{tid == 0} {
				Global\_cooperative\_memorize\_best\_worst\_sol($localbest\_thread$, $localworst\_thread$, $globalbest$, $globalworst$, $nv$, $nh$)\;
			}
			barrier()\;
			Local\_copy\_sols($globabest$, $globalworst$, $localbest$, $localworst$, $nv$, $nh$)\;
			Parallel\_update\_the\_population\_of\_sols($X$, $n$, $m$, $fv$, $localbest$, $localworst$, $nh$, $nv$)\;
			$iteration \leftarrow iteration + 1$\;
		}
	}
	Print\_the\_best\_solution\;
	\caption{Hierarchical hyper-population cooperative  parallel Jaya algorithm}
	\label{algo7}
\end{algorithm2e}
The pseudo-code of Algorithm~\ref{algo7} presents the overall structure of the HHCPJaya parallel computation. Initially, each worker thread initialises and evaluates a number $conf\_h$ of $nv$ vertical local subpopulations sequentially and independently on the population matrix $X$ and the fitness matrix $fv$, respectively (lines 3-4). In this initialisation phase, a different seed is employed for each local subpopulation, in order to increase the diversity of the subpopulations. During each iteration, each worker determines the local best and worst solutions for each of its local subpopulations on the respective subset of the population matrix $X$ independently, and these solutions are stored in the corresponding subsets of the matrices $localbest$ and $localworst$, respectively (line 6). Then, each worker updates its local subpopulations on the respective subsets of the matrix $X$ and the fitness matrix $fv$ based on the corresponding subsets of the matrices $localbest$ and $localworst$, according to Algorithm~\ref{algo5} (line 8). Following the local updating phase for the local subpopulations, all workers again perform the procedure $Local\_memorise\_best\_worst\_sol$, as we explained earlier (as seen in line 6). Lines 10 and 13 of Algorithm~\ref{algo7} enact the two-level hierarchical cooperative search scheme among the subpopulations. At the first level, each worker determines the local best and worst solutions from all of its local subpopulations. Then, these solutions are copied into a corresponding row of the two corresponding matrices $localbest\_thread$ and $localworst\_thread$ of size $threads \times bc$ in the shared memory, overwriting the previous solutions (line 10). At the second level, the master determines the global best and worst solutions from the corresponding matrices $localbest\_thread$ and $localworst\_thread$, and these global solutions are copied into vectors $globalbest$ and $globalworst$ of size $bc$ in the shared memory (line 13). Finally, each worker thread copies the global best and worst solutions in the corresponding shared matrices $localbest$ and $localworst$, and then follows the updating phase on each of its local subpopulations based on the global solutions (lines 15-16).

At the end of the parallel algorithm, each worker merges the local best solutions from its local subpopulations to form a complete solution with the original problem dimensions, and then calculates the fitness value based on the full solution. Finally, the master reports the optimal final result with the smallest fitness value from all workers (line 19).

The details of the parallel updating phase are presented in Algorithm~\ref{algo81}. Algorithms~\ref{algo7} and~\ref{algo81} are valid for any values of $n$, $m$, $nh$, $nv$, $conf\_h$, $conf\_v$, and $threads$. On the other hand, the procedures in lines 6, 9, 10, 13, and 15 of Algorithm~\ref{algo7} are simple, and their details are omitted owing to the space limitations.

\begin{algorithm2e}
	%\DontPrintSemicolon
	\KwIn{$X$: subpopulation matrix, $n$: population size, $m$: the number of variables, $fv$: fitness vector, $localbest, localworst$: the local best and worst solution vectors, $nh$: number of horizontal subpopulations, $nv$: number of vertical subpopulations}
	\KwOut{$X$: updated subpopulation matrix, $fv$: updated fitness matrix}
	$br \leftarrow n / nv$\;
	$bc \leftarrow m / nh$\;
	\For{$row \leftarrow 0$ \KwTo $nh/threads$}{
	\For{$col \leftarrow 0$ \KwTo $nv$}{
		\For{$i \leftarrow row * br$ \KwTo $(row + 1) * br$}{
		  \For{$j \leftarrow col * bc$ \KwTo $(col + 1) * bc$}{
			$X^{new}_{j}=X_{i,j} + r_{1}(localbest_{row,j}-|X_{i,j}|) - r_{2}(localworst_{row,j}-|X_{i,j}|)$\;
		}
		$fnew \leftarrow f(X^{new})$\;
		\If{$fnew < fv_{col,i}$}{
			\For{$j\leftarrow col*bc$ \KwTo $(col+1)*bc$}{
				$X_{i,j} \leftarrow X^{new}_{j}$\;
			}
			$fv_{col,i} \leftarrow fnew$\;
		}
	}	
  }
}
	\caption{Parallel\_update\_the\_population\_of\_solutions}
	\label{algo81}
\end{algorithm2e} 

Finally, the proposed decomposition strategy of the HHCPJaya algorithm is a general one, in the sense that the subpopulations can only be obtained by horizontal or vertical subpopulations of the original population matrix $X$ rather than block subpopulations, as discussed earlier. This means that the values of the parameters $nh$, $nv$, and $threads$ can be changed without changing the overall structure of the algorithm. Based on these parameters, the number and sizes of the subpopulations are adjusted.

\section{Experimental Results}
\label{results}
For the performance comparison of the proposed parallel algorithm, the computational experiments were organised into three aspects. The first involves a comparison of the parallel algorithm in terms of the solution quality. The second compares the performance of the parallel algorithm in terms of the parallel computation times and speedup. Finally, the third conducts an evaluation of the convergence ability of the parallel algorithm.

\subsection{Benchmark Functions}
These three types of experiments are performed on five representative benchmark functions of CEC2013 \cite{Li13benchmarkfunctions} with different characteristics, as listed in Table~\ref{functions}, which are specifically designed to benchmark large-scale optimisation algorithms. These functions can hardly get favorable results by current optimisation algorithms.

Function $f_{1}$ is Sphere function that is continuous, convex and unimodal. It is also simple and easy to be optimised. Function $f_{2}$ is Rastrigin function. It is highly multimodal and it also presents many local minima, regularly distributed through the large search space. Function $f_{3}$ is Ackley function that is continuous and multimodal in a generalized scalable version. Function $f_{4}$ is Griewank function. This function is strongly multimodal since the number of local optima increases with the dimensionality. However, this multimodality decreases as the size increases to higher dimensions. Function $f_{5}$ is Rosenbrock function which is continuous and convex. For higher dimensions, it turns into a multimodal function \cite{Shang:2006:NER:1118006.1118014}. The convergence to the global optimum of this function is difficult. 
	
The above functions with different properties (easy and difficult functions), are used to evaluate the quality of solutions and the convergence speed of the HHCPJaya algorithm. It is also interesting to see the parallel performance of the proposed algorithm for these functions when the number of variables scale. Furthermore, the difficult part of the multimodal functions ($f_{2}$, $f_{3}$, and $f_{5}$) is that an optimisation algorithm can easily be trapped in a local optimum on its way towards the global optimum. Hence, the hyper-population approach of the proposed HHCPJaya algorithm increases the diversity of the population and avoid premature convergence with satisfactory convergence speed. Another interesting point is that the optimisation of the multimodal functions are more challenging when the dimensionality of the problem increases.

\begin{sidewaystable}
\begin{tabular}{@{}llclc}
\hline
Name & Interval & C & Function & Global optimum \\ \hline
Sphere & $[-100,100]^{m}$ & US & $f_{1}(x) = \sum_{i=1}^m  x_i^2$ & $f_{1}(0,0,\dots)=0$\\
Rastrigin & $[-5, 5]^{m}$ & MS & $f_{2}(x) = 10 + \sum_{i=1}^m (x_i^2 -10cos(2\pi x_i))$ & $f_{2}(0,0,\dots)=0$ \\
Ackley & $[-32,32]^{m}$ & MN & $f_{3}(x) = -20 exp(-0.2 \sqrt{\frac{1}{m} \sum_{i=1}^m x_i^2}) - exp(\frac{1}{m} \sum_{i=1}^m cos(2\pi x_i)) + 20 + e$ &  $f_{3}(0,0,\dots)=0$\\
Griewank & $[-600,600]^{m}$ & MN & $f_{4}(x) = 1 + \frac{1}{4000} \sum_{i=1}^m x^2_i - \prod_{i=1}^m cos(\frac{x_i}{\sqrt{i}})$ & $f_{4}(0,0,\dots)=0$\\
Rosenbrock & $[-100,100]^{m}$ & UN & $f_{5}(x) = \sum_{i=1}^{m-1} 100(x_i^2 - x_{i+1})^2 + (1-x_i)^2$ & $f_{5}(1,1,\dots)=0$ \\
\hline
\end{tabular}
\caption{Numerical benchmark functions used for comparison (C: Characteristic, U: Unimodal, M: Multimodal, S: Separable, N:Non-Separable, m: Dimension)}
\label{functions}
%\end{table}
\end{sidewaystable}

\subsection{Experimental Environment}
The proposed parallel algorithm, HHCPJaya, is implemented in the C language using the OpenMP programming framework. The source code of the HHCPJaya algorithm can be found in the following link: \url{http://users.uom.gr/~pmichailidis/codehhcpjaya/}. The experiments are executed on a Dual Opteron 6128 CPU with eight processor cores (16 total cores), a 2.0 GHz clock speed, and 16 GB of memory, in a Ubuntu Linux 10.04 LTS environment. During all of the experiments, this machine was in exclusive use by this research project. To compile the multi-thread programs,
the Intel C/C++ compiler version 13.0, i.e., icc, icpc, is used, because this is a very widely used compiler. Finally, the double-precision floating-point format was used in the experiments.

\subsection{Performance Parameters}
The parameters that are affected by the performance of the HHCPJaya algorithm are the population size $n$, the number of variables $m$, the number of cores, and the parameters of the hyper-population approach, such as the number of subpopulations $k$ and the size of subpopulations $br \times bc$. The experiments were evaluated with a population size of 64 and 1024, numbers of variables of 512, 1024, and 2048, and numbers of cores from 1 to 16. As we discussed earlier, the number and sizes of the subpopulations are controlled by the number of horizontal subpopulations $nh$ and vertical subpopulations $nv$. The values of $nh$ and $nv$ depend on the values of the coefficients $conf\_h$ and $conf\_v$, the number of threads $threads$, and the values of the problem size $n$ and $m$. Table~\ref{nsubs} shows the values of $nh$, $nv$, $k$, $br$, and $bc$ that are computed using the equations~\ref{eqnh},~\ref{eqnv},~\ref{eqk},~\ref{eqbr}, and~\ref{eqbc}, respectively, with reference values $conf\_h=1$ and $conf\_v=1$ for each number of cores. For a certain values of $n$ and $m$ with all values of $conf\_h >1$ and $conf\_v>1$, the number of subpopulations $k$ is increased by $conf\_h \cdot conf\_v$ on the respective reference values of Table~\ref{nsubs}, and the sizes $br$ and $bc$ are decreased by $conf\_h$ and $conf\_v$ on the respective reference values of Table~\ref{nsubs}, respectively. For example, when the optimisation problem consists of $n=64$, $m=512$, $conf\_h=2$, and $conf\_v=2$, then the number of subpopulations $k$ is increased by four on the four cores, i.e, from 16 to 64, and the sizes of the subpopulations $br \times bc$ are decreased by two for both $br$ and $bc$, i.e., from $16\times 128$ to $8 \times 64$ on the four cores. Finally, the experiments were repeated 60 times, and the maximum number of function evaluations was chosen to be 128000.

%\begin{sidewaystable}
\begin{table}[h]
	\centering
	\begin{tabular}{l|lllll}
		\hline
		Problem sizes  &  \multicolumn{5}{c}{$n=64, m=512$}  \\ \hline
		Cores & $nh$ &	$nv$&	$k$ &	$br$	&b$c$ \\
		\hline
		1	&	1	& 1 &	1&	64 &	512\\
		2	&	2	& 2	& 4 & 32	& 256\\
		4	&	4 &	4 &	16 & 16 &	128\\
		8	&	8 &	8 &	64 & 	8&	64\\
		16 &		16 &	16 &	256 &	4	&32\\	
		& \multicolumn{5}{c}{$n=64, m=1024$}  \\ \hline
	1	&	1	& 1 &	1&	64 &	1024\\
	2	&	2	& 2	& 4 &	32	& 512\\
	4	&	4 &	4 &	16 &	16 &	256\\
	8	&	8 &	8 &	64 &	8 &	128\\
	16 &		16 &	16 &	256 &	4	&64\\
		& \multicolumn{5}{c}{$n=64, m=2048$}  \\ \hline
			1	&	1	& 1 &	1&	64 &	2048\\
			2	&	2	& 2	& 4 &	32	& 1024\\
			4	&	4 &	4 &	16 &	16 &	512\\
			8	&	8 &	8 &	64 &	8&	256\\
			16 &		16 &	16 &	256 &	4	&128\\
		\hline
	\end{tabular}
	\caption{Number and sizes of the subpopulations for several values of $n$ and $m$ with reference values $conf\_h=1$ and $conf\_v=1$}
	\label{nsubs}
	\end{table}
%\end{sidewaystable}

\subsection{Results for the Solution Quality Analysis}
The first set of experiments is conducted to check the quality of the solutions for the HHCPJaya algorithm. The solution quality can be defined as the distance of the obtained solution by the proposed parallel algorithm to the known global optimal solution of the benchmark function. The mean and standard deviation statistics of the function fitness values obtained by the HHCPJaya algorithm are presented in Table~\ref{tab1} for $m=512$, Table~\ref{tab2} for $m=1024$, and Table~\ref{tab3} for $m=2048$ for the population size $n=64$ and in Table~\ref{tab4} for $m=2048$ for the population size $n=1024$. In the tables, the best fitness results are depicted in bold face. Furthermore, three sets of results are displayed in each table. For the first set of results, we ran the algorithm for the reference values $conf\_h=1$ and $conf\_v=1$. In the second set of results, we tested the algorithm for several values of $conf\_h$ and $conf\_v$, and selected the values of $conf\_h$ and $conf\_v$ for which the algorithm reached the first global optimal solution in significantly less time. In the third set of results, we doubled the values of the parameters $conf\_h$ and $conf\_v$, based on the values of the parameters $conf\_h$ and $conf\_v$ of the second set of results. Furthermore, there are some asterisks present in the tables for the Rosenbrock function (i.e., the function $f_{5}$) on the 16 cores, because the subpopulations obtained for the selected values of $conf\_h$ and $conf\_v$ contain one variable, and the Rosenbrock function requires at least two variables to compute the fitness value. Moreover, statistical tests were conducted over the results of Tables~\ref{tab1} - ~\ref{tab4}
to better understand and support the conclusions. The two-sided Wilcoxon rank-sum test was used to check whether the differences between two algorithms, if any, are statistically significant. This test showed that the serial and HHCPJaya algorithms for the five benchmark functions were statistically different with a significance level of 0.05.

\begin{sidewaystable}
	\begin{tabular}{l|llll|llll|llll}
		\hline
		Function  &  \multicolumn{4}{c|}{$f_{1}$} &  \multicolumn{4}{c|}{$f_{2}$} &  \multicolumn{4}{c}{$f_{3}$}  \\ \hline
		Conficients & \multicolumn{4}{c|}{$conf\_h=1, conf\_v=1$} & \multicolumn{4}{c|}{$conf\_h=1, conf\_v=1$} & \multicolumn{4}{c}{$conf\_h=1, conf\_v=1$} \\ \hline
		Cores & Mean & StdDev & Time (s) & StdDev & Mean & StdDev & Time (s) & StdDev & Mean & StdDev & Time (s) & StdDev \\ \hline
		1	& 6.11E+04 &	6.85E+03 &	3.4338 &	0.0037 &	4.07E+03 &	2.57E+02 &	5.3109 &	0.0086&	1.74E+01 &	1.59E+00 &	5.2212 & 0.0084 \\
		2	& 1.69E+02	&6.62E+01 &	1.7531 & 0.0041 & 2.87E+03 & 4.39E+02 &	2.6962 &	0.0038	 &6.13E+00 &	8.15E-01 &	2.6611 & 0.0055 \\
		4	& 1.00E-06	&0	& 0.8973 &	0.0039	& 1.53E+03 &	2.33E+02 &	1.3691	& 0.0018 & 1.03E-01 &	3.15E-01 &	1.3643 &	0.0043\\
		8	& {\bf 0} &	0	 &0.4720 &	0.0060 &	{\bf 1.12E+03} &	7.46E+01 &	0.7111 &	0.0039 &	{\bf 5.03E+00} &	7.25E+00 &	0.7188 &	0.0072\\
		16	& {\bf 0}	& 0	 & 0.2715 &	0.0112&	1.81E+03	& 6.88E+01 &	0.3963	 & 0.0144 &	1.08E+01 & 7.69E+00 &	0.4073 &	0.0049\\
		\hline
		& \multicolumn{4}{c|}{$conf\_h=1, conf\_v=2$} & \multicolumn{4}{c|}{$conf\_h=1, conf\_v=16$} & \multicolumn{4}{c}{$conf\_h=1, conf\_v=4$} \\ \hline
		1 &	1.97E+03 &	4.68E+02 &	3.4545 &	0.0033 &	3.60E+03 &	8.36E+01 &	5.6617 &	0.0087 &	1.29E+00 &	3.52E+00 & 5.3588 & 0.0029\\ 
		2 &	1.87E-03 &	0 &	1.7693 &	0.0031 &	2.32E+03 &	4.72E+01 &	3.0075 &	0.0118 &	3.05E-01 &	9.25E-01 &	2.8048 &	0.0059\\
		4 & {\bf 0} &	0 &	0.9137 & 0.0022 & 8.41E+02 & 2.57E+01 & 1.6130 &	0.0194 &	1.06E+00 &	1.53E+00 &	1.4995 &	0.0057\\
		8 &	{\bf 0} &	0 &	0.4933 & 0.0063 & 3.61E+01 & 6.82E+00 &	0.9648 & 0.0036 & 1.77E-01 & 9.61E-01 &	0.8364 & 0.0055\\
		16 & {\bf 0} & 0 &	0.2958 &	0.0110 & {\bf 0 } &	0	& 0.6803 &	0.0153 &	{\bf 0} &	0 &	0.5041 &	0.0128\\
		\hline
		& \multicolumn{4}{c|}{$conf\_h=2, conf\_v=4$} & \multicolumn{4}{c|}{$conf\_h=2, conf\_v=32$} & \multicolumn{4}{c}{$conf\_h=2, conf\_v=8$} \\ \hline
		1 &	1.74E-03 &	5.12E-04 &	3.5037 &	0.0026 &	2.18E+03 &	4.57E+01 &	6.0586 &	0.0054 &	2.96E-01& 8.21E-01 & 5.5522 & 0.0058\\
		2  & {\bf 0} & 0	 & 1.8151 &	0.0033 &	7.16E+02 &	2.45E+01 &	3.2133 &	0.0602 &	5.36E-01 &	1.03E+00 &	2.9885 & 0.0054\\
		4 &	{\bf 0} &	0	& 0.9673 &	0.0034 &	7.57E+00 &	3.41E+00 &	1.9252	& 0.0109 &	1.02E-03 &	7.10E-03 &	1.6604 &	0.0070\\
		8 &	{\bf 0} &	0 &	0.5543 &	0.0036 &	{\bf 0} &	0 &	1.3263	& 0.0058 &	{\bf 0}	& 0 &	0.9761 &	0.0129\\
		16 & {\bf 	0} &	0 &	0.3434 &	0.0060 & {\bf 	0} &	0 &	1.1634 &	0.0198 & {\bf 0} &	0 &	0.6858 &0.0076\\
		\hline
	\end{tabular}
	\caption{Function fitness values of the solutions obtained by the HHCPJaya algorithm for the benchmark functions ($n=64, m=512$)}
	\label{tab1}
\end{sidewaystable}

\addtocounter{table}{-1}
\begin{sidewaystable}
	\begin{tabular}{l|llll|llll}
		\hline
		Function  &  \multicolumn{4}{c|}{$f_{4}$} &  \multicolumn{4}{c}{$f_{5}$}  \\ \hline
		Conficients & \multicolumn{4}{c|}{$conf\_h=1, conf\_v=1$} & \multicolumn{4}{c}{$conf\_h=1, conf\_v=1$}  \\ \hline
		Cores & Mean & StdDev & Time (s) & StdDev & Mean & StdDev & Time (s) & StdDev \\ \hline
		1 &	5.46E+02 &	5.76E+01 &	6.4695 &	0.0038 &	5.06E+09 &	1.44E+09 &	3.8061 &	0.0042\\
		2 &	1.96E+00 &	9.34E-01 &	3.2767&	0.0042 &	2.34E+09 &	3.24E+09 &	1.9405 &	0.0053\\
		4 &	{\bf 0} &	0 &	1.6652 &	0.0040	& 1.21E+09 &	3.14E+09 &	0.9924 &	0.0048\\
		8	 & {\bf 0} &	0 &	0.8607 &	0.0058 &	{\bf 2.66E+07} &	7.18E+07 &	0.5191 &	0.0039\\
		16	& {\bf 0} &	0 &	0.4718 &	0.0203 &	4.66E+09 &	1.78E+10 &	0.2976 &	0.0080\\		
		\hline
		& \multicolumn{4}{c|}{$conf\_h=1, conf\_v=2$} & \multicolumn{4}{c}{$conf\_h=1, conf\_v=16$}  \\ \hline
		1 &	1.93E+01 &	6.04E+00 &	6.5003 &	0.0031 &	3.73E+09 &	1.92E+10 &	4.1286 &	0.0072\\
		2 &	1.80E-05 &	6.00E-06 & 3.3052 &	0.0054 &	2.79E+06 &	2.16E+07 &	2.2515 &	0.0030\\
		4 & {\bf 0} &	0 &	1.6945 &	0.0035 &	6.10E+02 &	1.75E+03 &	1.1774 &	0.0111\\
		8 & {\bf 0} &	0 &	0.8914 &	0.0066 &	1.33E+03 &	3.43E+03 &	0.7162 &	0.0071\\
		16 & {\bf 0} & 0 &	0.4991 &	0.0095 &	{\bf 0} &	0 &	0.4983 &	0.0106\\		
		\hline
		& \multicolumn{4}{c|}{$conf\_h=2, conf\_v=4$} & \multicolumn{4}{c}{$conf\_h=2, conf\_v=32$}  \\ \hline
		1 &	1.70E-05 &	5.00E-06 &	6.5742 &	0.0028 &	2.24E+07 &	1.56E+08 &	4.4830 &	0.0048\\
		2	& {\bf 0} &	0	& 3.3766 &	0.0025 &	1.76E+02 &	9.68E+02 &	2.3413 &	0.0087 \\
		4 &	{\bf 0} &	0 &	1.7653 &	0.0079 &	3.85E+00 &	6.12E+00 &	1.4355 &	0.0127\\
		8 &	{\bf 0} &	0 &	0.9614 &	0.0070 &	{\bf 0} &	0 &	0.9709 &	0.0105\\
		16 & {\bf 0}	& 0	& 0.5562 &	0.0081 &	* &	* &	* &	* \\	
		\hline
	\end{tabular}
	\caption{(continued)}
\end{sidewaystable}

\begin{sidewaystable}
	\begin{tabular}{l|llll|llll|llll}
		\hline
		Function  &  \multicolumn{4}{c|}{$f_{1}$} &  \multicolumn{4}{c|}{$f_{2}$} &  \multicolumn{4}{c}{$f_{3}$}  \\ \hline
		Conficients & \multicolumn{4}{c|}{$conf\_h=1, conf\_v=1$} & \multicolumn{4}{c|}{$conf\_h=1, conf\_v=1$} & \multicolumn{4}{c}{$conf\_h=1, conf\_v=1$} \\ \hline
		Cores & Mean & StdDev & Time (s) & StdDev & Mean & StdDev & Time (s) & StdDev & Mean & StdDev & Time (s) & StdDev \\ \hline
		1 &	2.26E+05 &	2.06E+04 &	6.8595 &	0.0094 &	9.87E+03 &	3.33E+02 &	10.6117 &	0.0158 &	1.93E+01 &	1.65E+00& 	10.4118 &	0.0103\\
		2 &	2.63E+04 &	3.48E+03 &	3.4675 &	0.0083 &	7.34E+03 &	5.24E+02 &	5.3519 &	0.0096 &	1.30E+01&	5.71E-01&	5.2547 &	0.0097\\
		4 & 1.63E+00 &	5.40E-01 &	1.7698 &	0.0053 &	4.85E+03 &	4.96E+02 &	2.7107 &	0.0054 &	2.70E+00 &	3.26E-01 &	2.6789 &	0.0047\\
		8 &	{\bf 0} &	0 &	0.9172 &	0.0042 &	{\bf 3.65E+03} &	1.96E+02 &	1.3955 &	0.0117 &	{\bf 1.68E+00} &	5.08E+00 &	1.3851 &	0.0060\\
		16 & {\bf 	0} &	0 &	0.4943 &	0.0106 &	5.98E+03 &	1.39E+02 &	0.7401 &	0.0103 &	1.42E+01 &	7.17E+00 &	0.7454 &	0.0143\\	
		\hline
		& \multicolumn{4}{c|}{$conf\_h=1, conf\_v=2$} & \multicolumn{4}{c|}{$conf\_h=1, conf\_v=32$} & \multicolumn{4}{c}{$conf\_h=1, conf\_v=8$} \\ \hline
		1 &	6.33E+04 &	7.94E+03 &	6.8692 &	0.0080 &	7.36E+03 &	1.12E+02 &	11.3408 &	0.0179	& 1.60E+00 &	4.16E+00 &	10.7155 & 0.0071 \\
		2 &	4.96E+01 &	1.29E+01 &	3.4854 &	0.0089 &	4.66E+03 &	7.27E+01 &	6.0116&	0.0187 &	1.27E+00 &	2.03E+00 &	5.5704 &	0.0086\\
		4 &	4.00E-06 &	1.00E-06 &	1.7878 &	0.0044 &	1.71E+03 &	4.22E+01 &	3.2184	&0.0076 &	1.59E+00 &	1.67E+00 &	3.0035 &	0.0039\\
		8 & {\bf 	0} &	0 &	0.9310 &	0.0033 &	8.13E+01 &	1.37E+01 &1.9338 &	0.0114 &	9.65E-02 &	7.48E-01 &	1.6714 &	0.0085\\
		16 & {\bf 	0} &	0 &	0.5172 &	0.0107 &	{\bf 0} &	0 &	1.3397 &	0.0104 &	{\bf 0} &	0 &	0.9904 &	0.0326\\	
		\hline
		& \multicolumn{4}{c|}{$conf\_h=2, conf\_v=4$} & \multicolumn{4}{c|}{$conf\_h=2, conf\_v=64$} & \multicolumn{4}{c}{$conf\_h=2, conf\_v=16$} \\ \hline
		1 &	4.83E+01 &	1.32E+01 &	6.9397 &	0.0124 &	4.40E+03 &	6.58E+01 &	12.151	& 0.014	 & 9.49E-01 &	1.23E+00 &	11.1095 &	0.0160\\
		2	& 4.00E-06 &	1.00E-06 &	3.5554 &	0.0074 &	1.45E+03 &	3.99E+01 &	6.443 &	0.030 &	1.38E+00 &	1.27E+00 &	5.9903 &	0.0070\\
		4 &	{\bf 0} &	0 &	1.8441 &	0.0046 &	1.85E+01 &	5.50E+00 &	3.853 &	0.008 &	1.49E-01 &	8.11E-01 &	3.3197 &	0.0112\\
		8	& {\bf 0} & 	0 &	1.0053 &	0.0067 &	{\bf 0} &	0 &	2.655 &	0.012 &	{\bf 0}	& 0 &	1.9639 &	0.0199\\
		16	& {\bf 0} & 	0 &	0.5887 &	0.0111 & {\bf 0} &	0 &	2.301 &	0.015 &	{\bf 0} &	 0 &	1.3607 &	0.0089\\		
		\hline
	\end{tabular}
	\caption{Function fitness values of the solutions obtained by the HHCPJaya algorithm for the benchmark functions ($n=64, m=1024$)}
	\label{tab2}
\end{sidewaystable}

\addtocounter{table}{-1}
\begin{sidewaystable}
	\begin{tabular}{l|llll|llll}
		\hline
		Function  &  \multicolumn{4}{c|}{$f_{4}$} &  \multicolumn{4}{c}{$f_{5}$}  \\ \hline
		Conficients & \multicolumn{4}{c|}{$conf\_h=1, conf\_v=1$} & \multicolumn{4}{c}{$conf\_h=1, conf\_v=1$}  \\ \hline
		Cores & Mean & StdDev & Time (s) & StdDev & Mean & StdDev & Time (s) & StdDev \\ \hline
		1 &	2.04E+03 &	1.86E+02 &	12.9233 &	0.0073 &	2.39E+10  &	5.65E+09 &	7.6021 &	0.0074\\
		2	& 2.38E+02 &	3.45E+01 &	6.5038 &	0.0091 &	2.33E+09 &	2.88E+09 &	3.8383 &	0.0080\\
		4	& 1.55E-02 &	4.65E-03 &	3.2916 &	0.0054 &	2.82E+09 &	6.24E+09 &	1.9562 &	0.0044\\
		8	& {\bf 0} &	0 &	1.6862 &	0.0032 &	{\bf 8.04E+08} &	2.78E+09 &	1.0127 &	0.0066\\
		16	& {\bf 0} &	0 &	0.8834 &	0.0056 &	1.15E+10 &	2.86E+10 &	0.5472 &	0.0158\\		
		\hline
		& \multicolumn{4}{c|}{$conf\_h=1, conf\_v=2$} & \multicolumn{4}{c}{$conf\_h=1, conf\_v=32$}  \\ \hline
		1 &	5.73E+02 &	6.93E+01 &	12.9416 &	0.0082 &	7.07E+08 &	4.71E+09 &	8.2738 &	0.0145\\
		2 &	4.37E-01 &	1.11E-01 &	6.5300 &	0.0081 &	2.79E+06 &	1.74E+07 &	4.5017 &	0.0063\\
		4 & {\bf 0} &	0 &	3.3238 &	0.0048 &	4.70E+05 &	1.75E+06 &	2.3603 &	0.0074\\
		8 &	{\bf 0} &	0 &	1.7142 &	0.0052 &	1.69E+04	& 4.86E+04 &	1.4418 &	0.0031\\
		16 & {\bf 0} &	0 &	0.9153 &	0.0173 &	{\bf 0} &	0 &	0.9831 &	0.0114\\		
		\hline
		& \multicolumn{4}{c|}{$conf\_h=2, conf\_v=4$} & \multicolumn{4}{c}{$conf\_h=2, conf\_v=64$}  \\ \hline
		1 &	4.37E-01 &	1.29E-01 &	13.034 &	0.010 &	2.58E+06 &	2.00E+07 &	8.9867 &	0.0105\\
		2	& {\bf 0} &	0 &	6.627 &	0.007 &	3.25E+02 &	9.60E+02 &	4.7166 &	0.0130\\
		4	& {\bf 0} &	0	 & 3.408	& 0.007	 & 2.14E+01	& 3.02E+01 &	2.8782 &	0.0124\\
		8	& {\bf 0} & 	0 &	1.799 &	0.005 & {\bf	0} &	0 &	1.9407 &	0.0062\\
		16	& {\bf 0} &	0 &	0.965 &	0.015 &	*&	*	&*&	*\\	
		\hline
	\end{tabular}
	\caption{(continued)}
\end{sidewaystable}

\begin{sidewaystable}
	\begin{tabular}{l|llll|llll|llll}
		\hline
		Function  &  \multicolumn{4}{c|}{$f_{1}$} &  \multicolumn{4}{c|}{$f_{2}$} &  \multicolumn{4}{c}{$f_{3}$}  \\ \hline
		Conficients & \multicolumn{4}{c|}{$conf\_h=1, conf\_v=1$} & \multicolumn{4}{c|}{$conf\_h=1, conf\_v=1$} & \multicolumn{4}{c}{$conf\_h=1, conf\_v=1$} \\ \hline
		Cores & Mean & StdDev & Time (s) & StdDev & Mean & StdDev & Time (s) & StdDev & Mean & StdDev & Time (s) & StdDev \\ \hline
		1	& 5.61E+05 &	4.61E+04 &	13.84073 &	0.05219 &	2.11E+04 &	4.18E+02 &	21.30756 &	0.02647 &	1.99E+01 &	1.29E+00	& 20.88399 &	0.02870\\
		2	& 1.85E+05 &	1.15E+04 &	6.95908 &	0.02608 &	1.78E+04 &	8.24E+02 &	10.70139 & 0.02675 & 1.47E+01 & 2.67E-01 &	10.51463 &	0.01507\\
		4	& 5.40E+03 &	1.24E+03 &	3.52031 &	0.01395 &	1.47E+04 &	7.47E+02 &	5.39681 &	0.01574	& 7.95E+00 &	4.80E-01 &	5.32906 &	0.00944\\
		8 &	{\bf 0} &	0 &	1.81496 &	0.01233 &	{\bf 1.18E+04} &	4.14E+02 &	2.76200 &	0.01143 &	{\bf 9.11E-01} &	2.24E+00	& 2.72051 &	0.01336\\
		16 & {\bf 0} &	0 &	0.96079 &	0.00966 &	1.59E+04 &	2.08E+02 &	1.45668 &	0.02760 &	7.64E+00 &	9.11E+00 &	1.43381 &	0.01479\\		
		\hline
		& \multicolumn{4}{c|}{$conf\_h=1, conf\_v=2$} & \multicolumn{4}{c|}{$conf\_h=1, conf\_v=64$} & \multicolumn{4}{c}{$conf\_h=1, conf\_v=16$} \\ \hline
		1 &	3.14E+05 &	2.71E+05 &	13.78548 &	0.06830 &	1.48E+04 &	1.77E+02 &	22.73712 &	0.02746 &	3.12E+00 &	6.27E+00 &	21.46736 &	0.02457 \\
		2 &	1.70E+04 & 2.79E+03 &	6.94441 &	0.02365 &	9.35E+03 &	1.11E+02 &	12.03225 &	0.02347 &	2.08E+00 &	2.15E+00 &	11.14450 &	0.01771\\
		4 &	1.16E+00 &	2.22E-01 &	3.52780 &	0.01831 &	3.47E+03 &	6.54E+01 &	6.44818 &	0.05337 &	2.94E+00 &	1.52E+00 &	6.00760 &	0.00823\\
		8	& {\bf 0} &	0 &	1.82616 &	0.01181 &	1.71E+02 &	2.57E+01 &	3.87772 &	0.00575 &	1.21E+00 &	2.34E+00 &	3.34342 &	0.01647\\
		16 & {\bf 0} &	0 &	0.97086 &	0.01063 &	{\bf 0} &	0	 & 2.68973 &	0.01783 & {\bf 	0} &	0 &	1.96245 &	0.01290\\	
		\hline
		& \multicolumn{4}{c|}{$conf\_h=2, conf\_v=4$} & \multicolumn{4}{c|}{$conf\_h=2, conf\_v=128$} & \multicolumn{4}{c}{$conf\_h=2, conf\_v=32$} \\ \hline
		1 &	1.69E+04 &	2.53E+03 &	14.01392 &	0.01521	& 8.81E+03 & 1.08E+02 &	24.11894 &	0.04600	& 1.47E+00 & 1.26E+00 &	22.32107 &	0.02960\\
		2 & 1.15E+00 &	2.53E-01 &	7.12187 &	0.01312	& 2.92E+03 & 4.88E+01 &	12.96410 & 0.16783 &	2.02E+00 &	1.01E+00	& 12.13728 &	0.14421\\
		4 &	4.70E-05 &	8.00E-06 &	3.65527 &	0.01154 &	4.32E+01 & 1.10E+01 &	7.80367 &	0.05818 &	1.01E+00 &	1.84E+00 &	6,71330 &	0.05015\\
		8	& {\bf 0} &	0 &	1.93276 &	0.05790 &	{\bf 0} &	0 &	5.33692 &	0.03348	& {\bf 0} &	0 &	3.95968	 & 0.07246\\
		16	& {\bf 0 } &	0	 & 1.07153 &	0.02071 &	{\bf 0} &	0 &	4.69431 &	0.06049 &	{\bf 0} &	0 &	2.85839 &	0.01243\\	
		\hline
	\end{tabular}
	\caption{Function fitness values of the solutions obtained by the HHCPJaya algorithm for the benchmark functions ($n=64, m=2048$)}
	\label{tab3}
\end{sidewaystable}

\addtocounter{table}{-1}
\begin{sidewaystable}
	\begin{tabular}{l|llll|llll}
		\hline
		Function  &  \multicolumn{4}{c|}{$f_{4}$} &  \multicolumn{4}{c}{$f_{5}$}  \\ \hline
		Conficients & \multicolumn{4}{c|}{$conf\_h=1, conf\_v=1$} & \multicolumn{4}{c}{$conf\_h=1, conf\_v=1$}  \\ \hline
		Cores & Mean & StdDev & Time (s) & StdDev & Mean & StdDev & Time (s) & StdDev \\ \hline
		1	& 5.05E+03 &	4.15E+02 &	25.9322 &	0.0257 &	6.49E+10 &	4.31E+10 &	15.2821 &	0.0287\\
		2 &	1.66E+03 &	1.15E+02 &	13.0147 &	0.0199 &	7.80E+09	& 1.89E+09 &	7.7033 &	0.0150\\
		4 &	5.20E+01 &	1.18E+01 &	6.5515 &	0.0135 &	{\bf 3.01E+09} &	5.83E+09 &	3.8929 &	0.0148\\
		8	& {\bf 0} &	0 &	3.3379 &	0.0162 &	8.81E+09 &	1.35E+10 &	1.9999 &	0.0100\\
		16 & {\bf 	0} &	0 &	1.7289 &	0.0088 &	7.07E+09 &	2.53E+10 &	1.0634 &	0.0116\\	
		\hline
		& \multicolumn{4}{c|}{$conf\_h=1, conf\_v=2$} & \multicolumn{4}{c}{$conf\_h=1, conf\_v=64$}  \\ \hline
		1 &	2.83E+03 &	2.41E+02 &	25.9013	&0.0314	 & 8.89E+09 &	3.26E+16 &	16.5771 &	0.0300\\
		2 &	1.54E+02 &	2.52E+01 &	13.0149 &	0.0152 &	4.55E+13 &	3.10E+08 &	9.0140 &	0.0127\\
		4 &	1.05E-02 &	2.24E-03 &	6.5741 &	0.0147 &	2.97E+07 &	1.93E+08 &	4.7289 &	0,0170\\
		8 & {\bf 0} &	0 &	3.3638 &	0.0100 &	1.53E+05 &	2.15E+05 &	2.9044 &	0.0126\\
		16 & {\bf 0}	& 0 & 1.7558 & 0.0247 &	{\bf 0} &	0 &	1.9804 &	0.0138\\	
		\hline
		& \multicolumn{4}{c|}{$conf\_h=2, conf\_v=4$} & \multicolumn{4}{c}{$conf\_h=2, conf\_v=128$}  \\ \hline
		1 &	1.49E+02 &	2.43E+01 &	26.1174 &	0.0144 &	2.39E+06 &	1.72E+07 &	18.1091	 & 0.1886\\
		2 & 1.11E-02 & 2.69E-03 &	13.2272 &	0.2305 &	1.07E+03 &	3.04E+03 &	9.5271 &	0.1824\\
		4 &	{\bf 0} &	0 &	6.7201 &	0.0122	 &1.76E+02 &	3.60E+02 &	5.8674	& 0.0563\\
		8 &	 {\bf 0} &	0 &	3.5025 &	0.0875 &	{\bf 0} &	0 &	3.9181 &	0.0573\\
		16 & {\bf 0} &	0 &	1.8627 &	0.0163 &	* & 	* &	* &	* \\	
		\hline
	\end{tabular}
	\caption{(continued)}
\end{sidewaystable}

\begin{sidewaystable}
	\begin{tabular}{l|llll|llll|llll}
		\hline
		Function  &  \multicolumn{4}{c|}{$f_{1}$} &  \multicolumn{4}{c|}{$f_{2}$} &  \multicolumn{4}{c}{$f_{3}$}  \\ \hline
		Conficients & \multicolumn{4}{c|}{$conf\_h=1, conf\_v=1$} & \multicolumn{4}{c|}{$conf\_h=1, conf\_v=1$} & \multicolumn{4}{c}{$conf\_h=1, conf\_v=1$} \\ \hline
		Cores & Mean & StdDev & Time (s) & StdDev & Mean & StdDev & Time (s) & StdDev & Mean & StdDev & Time (s) & StdDev \\ \hline
		1	&	1.14E+06	&	4.30E+04	&	218.28532	&	1.44491	&	2.45E+04	&	3.64E+02	&	337.89980	&	0.93311		&	2.06E+01	&	2.57E-01	&	438.59007	&	3.68539\\
		2	&	6.47E+05	&	1.51E+04	&	110.25865	&	1.69444		&	2.24E+04	&	1.98E+02	&	170.55136	&	2.37723		&	2.03E+01	&	1.98E-01	&	219.35105	&	0.45795\\
		4	&	2.66E+05	&	6.73E+03	&	55.77359	&	0.80245		&	2.10E+04	&	1.70E+02	&	85.23827	&	0.49353		&	1.74E+01	&	1.08E+00	&	109.65898	&	0.11070\\
		8	&	3.18E+04	&	1.13E+03	&	27.97297	&	0.17685	&	1.98E+04	&	1.81E+02	&	43.34910	&	1.06171	&	9.78E+00	&	4.12E+00	&	55.35419	&	0.07013\\
		16	&	{\bf 2.10E+02}	&	7.42E+00	&	14.22020	&	0.19386		&	{\bf 1.78E+04}	&	2.40E+02	&	22.17339	&	0.94941	&	{\bf 7.41E+00}	&	3.16E+00	&	28.60177	&	0.44388\\				
		\hline
		& \multicolumn{4}{c|}{$conf\_h=2, conf\_v=4$} & \multicolumn{4}{c|}{$conf\_h=1, conf\_v=64$} & \multicolumn{4}{c}{$conf\_h=4, conf\_v=16$} \\ \hline
		1	&	6.19E+05	&	1.51E+04	&	222.28204	&	2.15298		&	1.67E+04	&	1.19E+02	&	376.98010	&	4.05662		&	1.87E+01	&	2.77E+00	&	447.93160	&	1.94359\\
		2	&	1.94E+05	&	5.67E+03	&	112.15885	&	0.85532		&	9.64E+03	&	7.88E+01	&	208.13558	&	2.40194		&	3.37E+00	&	4.09E-02	&	230.20457	&	1.25605\\
		4	&	7.33E+03	&	2.45E+02	&	56.28904	&	0.21313		&	3.34E+03	&	4.46E+01	&	100.62291	&	0.21986		&	9.39E-04	&	2.80E-05	&	120.95994	&	0.58979\\
		8	&	7.51E-01	&	2.53E-02	&	28.80661	&	0.37405		&	2.31E+02	&	1.15E+01	&	58.99824	&	0.11842		&	{\bf 0}	&	0	&	65.35867	&	0.30434\\
		16	&	{\bf 0}	&	0	&	15.50264	&	0.29424		&	{\bf 0}	&	0	&	41.70571	&	0.74590		&	{\bf 0}	&	0	&	37.64871	&	0.78423\\		
		\hline
		& \multicolumn{4}{c|}{$conf\_h=4, conf\_v=8$} & \multicolumn{4}{c|}{$conf\_h=2, conf\_v=128$} & \multicolumn{4}{c}{$conf\_h=8, conf\_v=32$} \\ \hline
		1	&	1.92E+05	&	4.59E+03	&	222.31140	&	1.29013	&	9.26E+03	&	8.70E+01	&	408.99803	&	3.43165			& 3.28E+00	&	3.74E-02	&	456.90438	&	0.21971\\
		2	&	7.09E+03	&	2.17E+02	&	113.19666	&	1.33131	&	3.03E+03	&	4.61E+01	&	204.74997	&	2.08642 &	8.27E-04	&	2.20E-05	&	242.45926	&	0.29105\\
		4	&	7.10E-01	&	2.15E-02	&	57.48221	&	0.75616	&	1.46E+02	&	4.86E+00	&	117.85297	&	1.12337	&	{\bf 0}	&	0	&	130.68160	&	0.48205\\
		8	&	{\bf 0}	&	0	&	30.43136	&	0.41884	& {\bf 0}	&	0	&	84.15657	&	0.74864 & {\bf 0}	&	0	&	73.38569	&	0.87989\\
		16	&	{\bf 0}	&	0	&	17.18475	&	0.32153	& {\bf 0}	&	0	&	73.24125	&	1.12256 & {\bf 0}	&	0	&	48.20241	&	1.22272\\
		\hline
	\end{tabular}
	\caption{Function fitness values of the solutions obtained by the HHCPJaya algorithm for the benchmark functions ($n=1024, m=2048$)}
	\label{tab4}
\end{sidewaystable}

\addtocounter{table}{-1}
\begin{sidewaystable}
	\begin{tabular}{l|llll|llll}
		\hline
		Function  &  \multicolumn{4}{c|}{$f_{4}$} &  \multicolumn{4}{c}{$f_{5}$}  \\ \hline
		Conficients & \multicolumn{4}{c|}{$conf\_h=1, conf\_v=1$} & \multicolumn{4}{c}{$conf\_h=1, conf\_v=1$}  \\ \hline
		Cores & Mean & StdDev & Time (s) & StdDev & Mean & StdDev & Time (s) & StdDev \\ \hline
		1 &	1.03E+04	&	3.87E+02	&	412.2185	&	3.7701		&	3.19E+11	&	2.25E+10	&	241.8436	&	1.1427\\
		2 &	5.83E+03	&	1.40E+02	&	206.5835	&	0.2519	&	1.01E+11	&	1.01E+10	&	122.3208	&	1.8210\\
		4 &	2.39E+03	&	5.17E+01	&	103.3615	&	0.1435	&	1.47E+10	&	9.63E+08	&	60.9696	&	0.1078\\
		8 &	2.87E+02	&	8.85E+00	&	52.0800	&	0.0659	&	{\bf 1.30E+09}	&	1.09E+09	&	30.8294	&	0.0634\\
		16 &	{\bf 1.91E+00}	&	8.08E-02	&	26.8579	&	0.6831	&	1.66E+09	&	4.73E+09	&	15.9250	&	0.2376\\		
		\hline
		& \multicolumn{4}{c|}{$conf\_h=4, conf\_v=4$} & \multicolumn{4}{c}{$conf\_h=1, conf\_v=64$}  \\ \hline
		1 &	2.36E+03	&	6.26E+01	&	414.6486	&	1.7138		&	1.08E+11	&	5.57E+10	&	278.4745	&	4.2088\\
		2 &	2.74E+02	&	9.02E+00	&	209.2265	&	1.0320	&	1.32E+04	&	2.56E+02	&	159.5720	&	2.4773\\
		4 &	1.77E+00	&	7.62E-02	&	105.5063	&	0.2080	&	2.03E+04	&	6.09E+02	&	72.9918	&	0.8800\\
		8 &	{\bf 0}	&	0	&	54.0629	&	0.4543	&	1.16E+03	&	1.51E+02	&	43.3506	&	0.6899\\
		16 & {\bf 0}	&	0	&	28.8639	&	0.9582	&	{\bf 0}	&	0	&	28.69000	&	0.6601\\		
		\hline
		& \multicolumn{4}{c|}{$conf\_h=8, conf\_v=8$} & \multicolumn{4}{c}{$conf\_h=2, conf\_v=128$}  \\ \hline
		1	&2.73E+02	&	9.75E+00	&	414.2771	&	0.1897	&	1.33E+04	&	2.27E+02	&	310.4639	&	2.6727\\
		2	&1.72E+00	&	7.74E-02	&	211.0481	&	0.7131	&	1.77E+04	&	5.28E+02	&	159.5952	&	1.4022\\
		4	&{\bf 0}	&	0	&	107.6781	&	0.6373	&	2.18E+02	&	1.85E+01	&	85.3504	&	0.8298\\
		8	& {\bf 0}	&	0	&	57.3821	&	1.5307	&	{\bf 0}	&	0	&	59.5263	&	0.6509\\
		16	& {\bf 0}	&	0	&	31.3656	&	0.8163	&	*	&	*	&	*	&	*\\
		
		\hline
	\end{tabular}
	\caption{(continued)}
\end{sidewaystable}

From the results, it is clear that for all benchmark functions and any values of $conf\_h$ and $conf\_v$ the proposed HHCPJaya algorithm improves the quality of solutions as the number of cores increases. This fact is down to the hyper-population approach, i.e., more subpopulations are used when the number of cores increases. However, the results show that the highest solution quality of the functions $f_{2}$, $f_{3}$ and $f_{5}$ for $conf\_h = 1$ and $conf\_v=1$ is achieved when the number of cores is 8. This is due to the nature of complex functions and the diversity of the subpopulations. It is possible the specific diversity of the subpopulations cannot able enough to improve further the solution quality. This phenomenon is disappeared for increasing values of $conf\_h$ and $conf\_v$ as we can observe in the experimental results.

For each benchmark function, we also observe that the proposed parallel algorithm produces the best results when the values of the parameters $conf\_h$ and $conf\_v$ are increased. This fact means that the algorithm further improves the quality of solutions as the number of cores and subpopulations is increased. It can be seen that the maximum number of small subpopulations ($br$ and $bc$) produces the best results in the experiments. For example, the solution obtained by the parallel algorithm for the Ackley function with $m=512$ improves when the pair of parameters $(conf\_h, conf\_v)$ increases from (1,1) to (1,4) on eight cores. The improved solution occurs because the number of subpopulations increases from 64 to 256 on eight cores and the size of the subpopulations decreases from $8 \times 64$ to $8 \times 16$ on eight cores. Finally, the solution is further improved and reaches the optimal solution when the pair of parameters is (2,8) on eight cores, because the number of subpopulations is significant (i.e., 1024) and the size of subpopulations is sufficiently small (i.e., $4 \times 8$) on eight cores.

Increasing the values of the parameters $conf\_h$ and $conf\_v$ also increases the number of small subpopulations. Note that an appropriate choice for the values of the parameters $conf\_h$ and $conf\_v$ depends on the problem size (i.e., the population size and the number of variables). When the size of the population and the dimensions of the problem are large, we select large values for $conf\_h$ and $conf\_v$. Otherwise, we select small values for $conf\_h$ and $conf\_v$. In the experiments presented here, we selected small and medium values for the parameter $conf\_h$ and medium/large values for the parameter $conf\_v$, because the population size is small and large and the dimensions of the problem are large.  However, by further increasing the value of the $conf\_h$ for a population of small or large size while maintaining the large value of $conf\_v$ for large dimensions, the HHCPJaya algorithm can quickly reach the optimum solution on a small number of cores, but not for a large number of cores. 
 
The improvement in the performance of the parallel algorithm may vary depending on the function. For instance, it is evident from the results that the convergence rate of the algorithm for complex functions, such as the Rastrigin, Ackley and Rosenbrock functions, is high when the values of the parameters $conf\_h$ and $conf\_v$ are large (i.e., the number of subpopulations is sufficiently large). This is down to the two-level hierarchical cooperative search scheme among the subpopulations and the local updating phase based on the global solutions. As such, significantly improving the success rate of the HHCPJaya algorithm for these complex functions would require additional generations or function evaluations to reach the global optimal solution.

Finally, the results show that the most cases the HHCPJaya algorithm with small population size $n=64$ seem to produce better solutions quality than the HHCPJaya with $n=1024$ for $conf\_h=1$ and $conf\_v=1$. This fact is due to a large population size could reduce exploration capabilities of a population based optimizer by increasing its structural bias \cite{KONONOVA2015468}.

\subsection{Results for the Parallel Time and Speedup Analysis}
\begin{figure}
	\subfigure[$f_{1}$]{
		\includegraphics[width=0.60\textwidth]{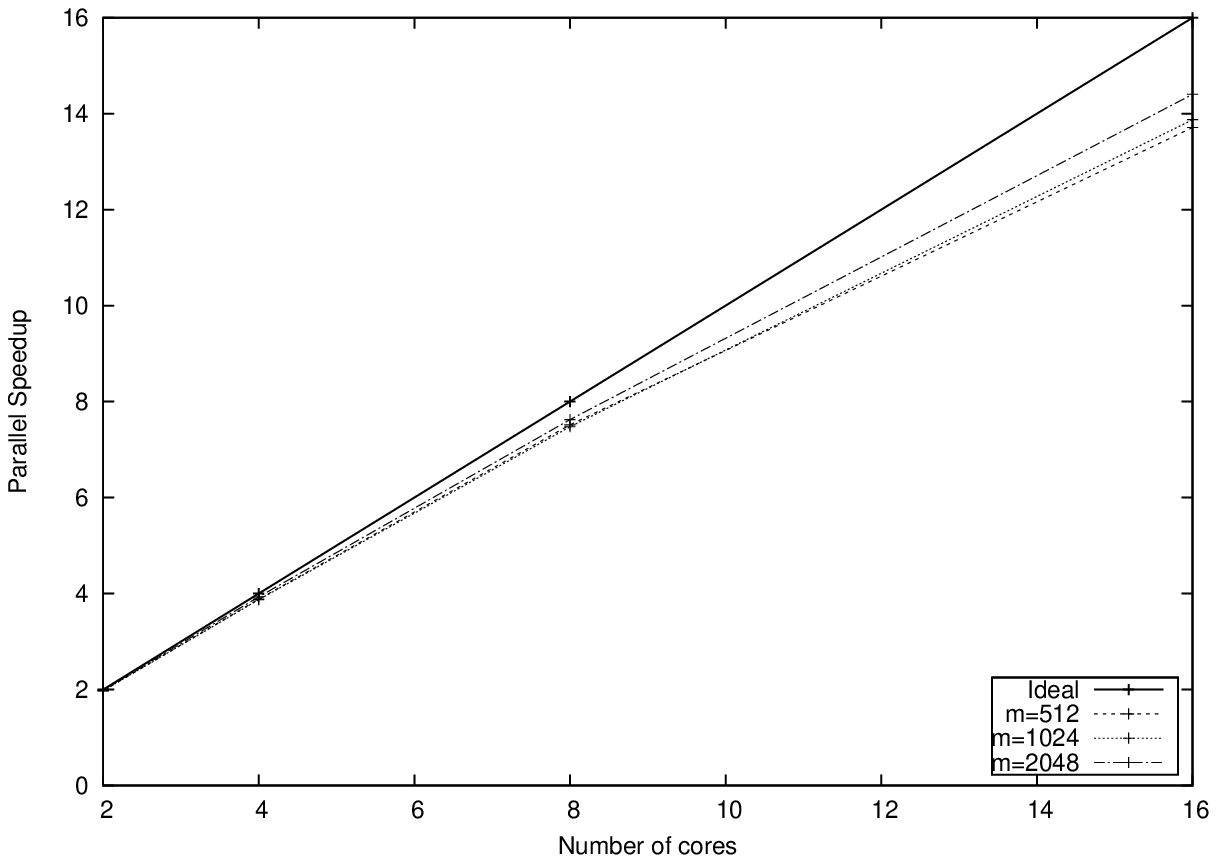}
	}
	\subfigure[$f_{2}$]{
		\includegraphics[width=0.60\textwidth]{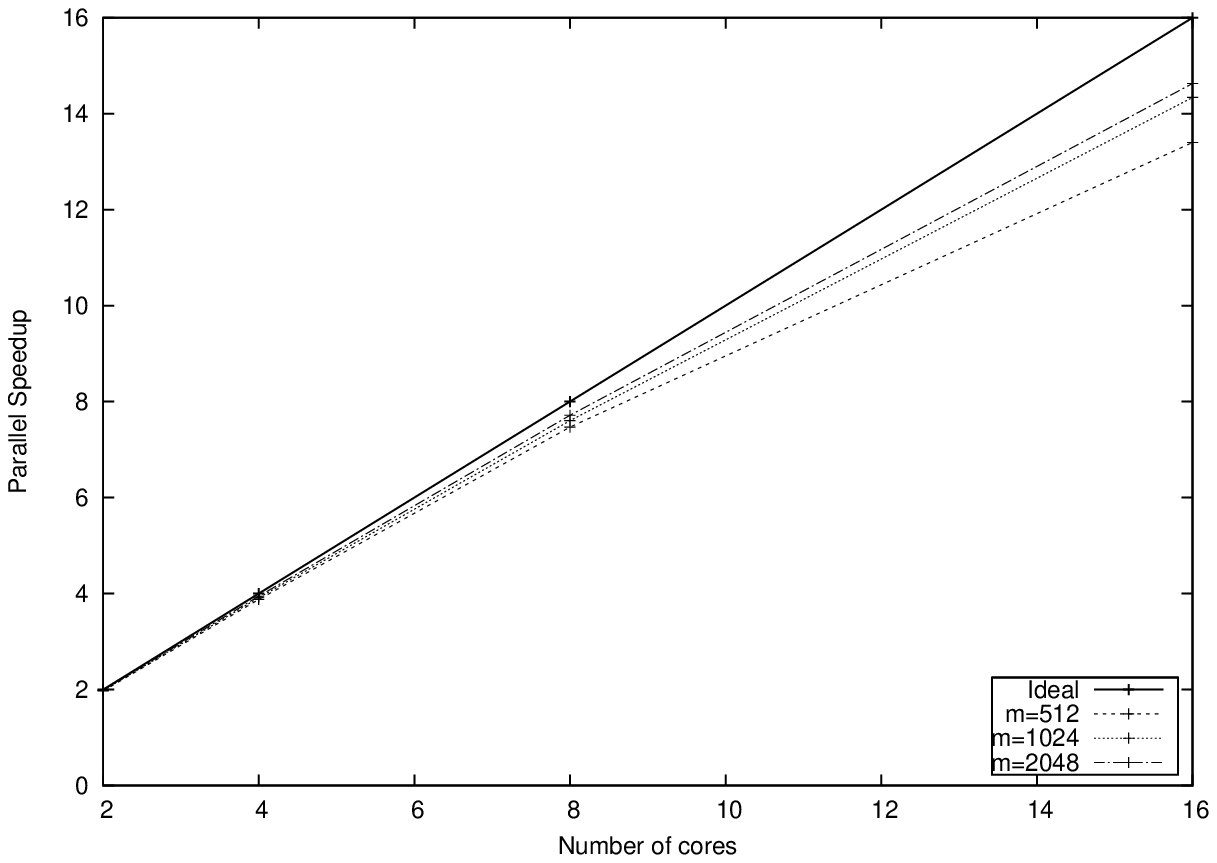}
	}
	\hfill
	\subfigure[$f_{3}$]{
		\includegraphics[width=0.60\textwidth]{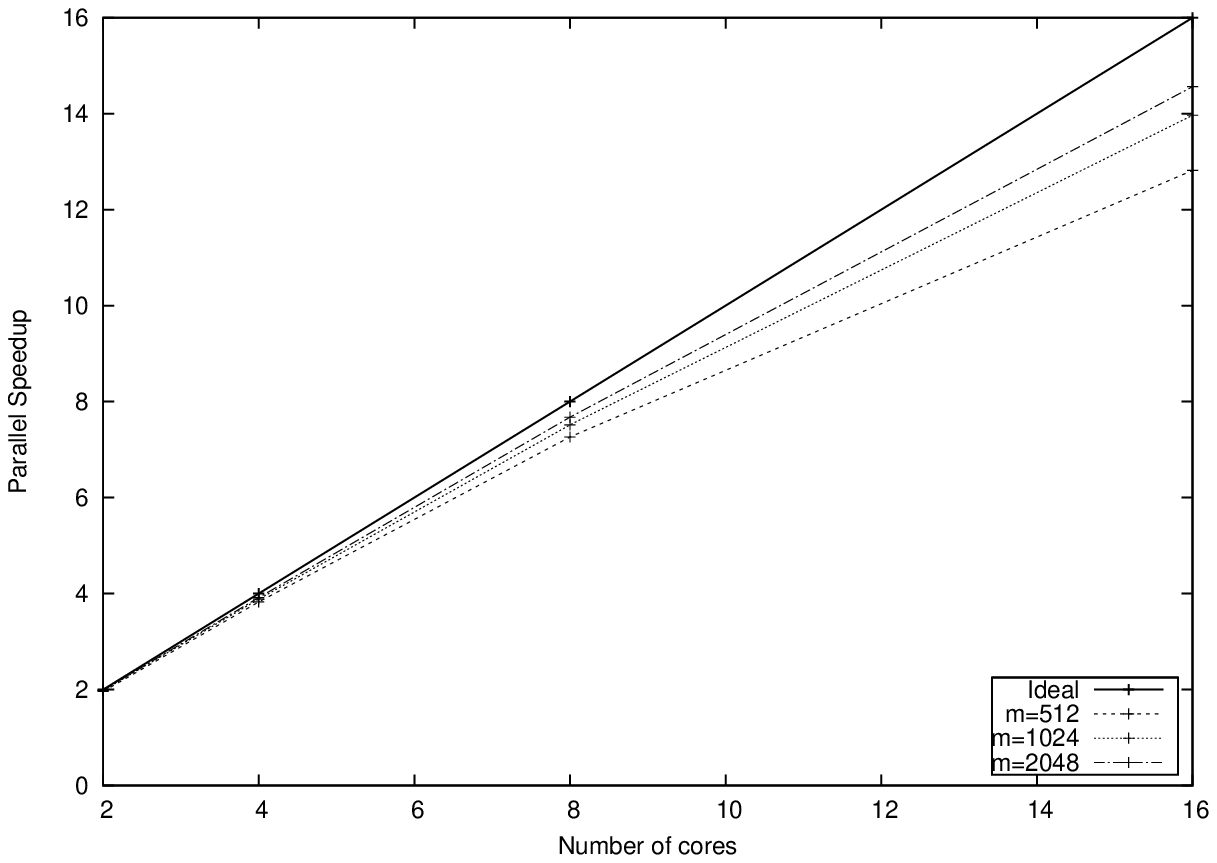}
	}
	\subfigure[$f_{4}$]{
		\includegraphics[width=0.60\textwidth]{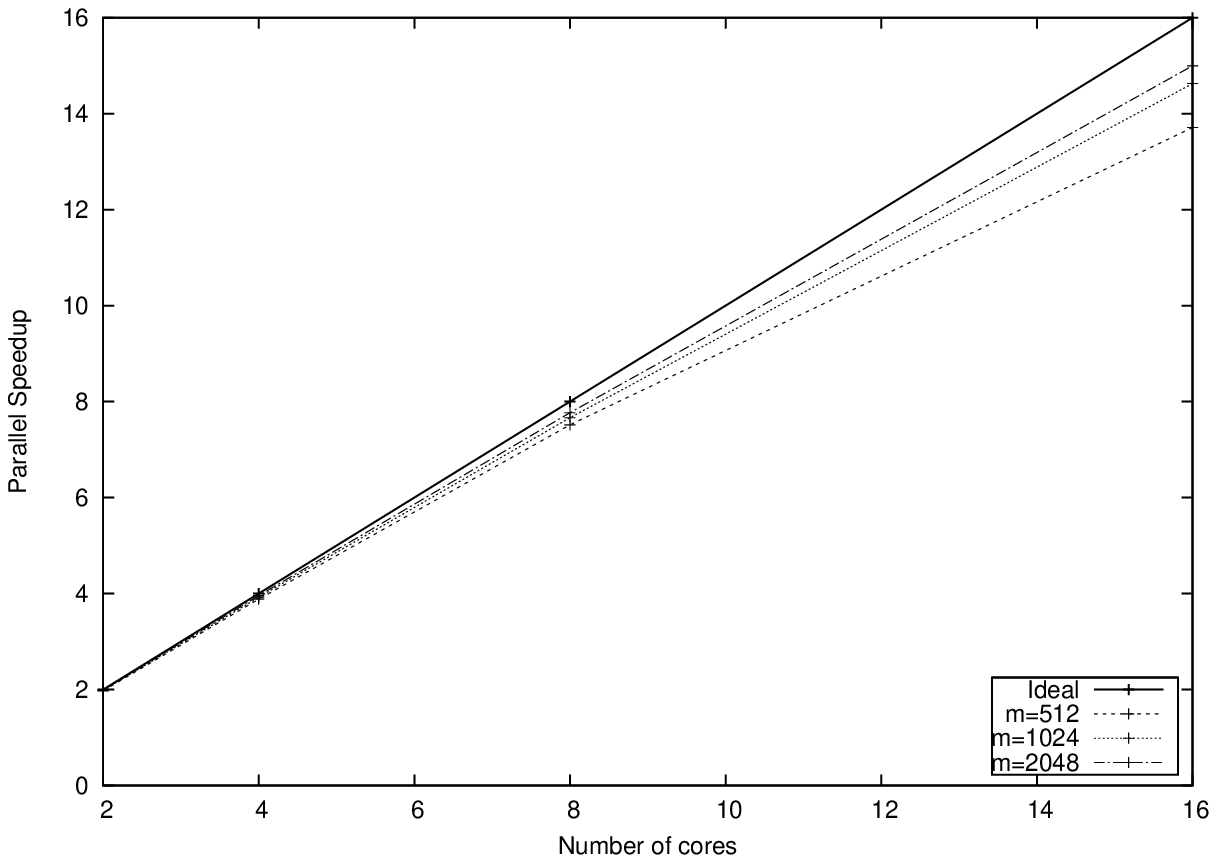}
	}
	\hfill
	\subfigure[$f_{5}$]{
		\includegraphics[width=0.60\textwidth]{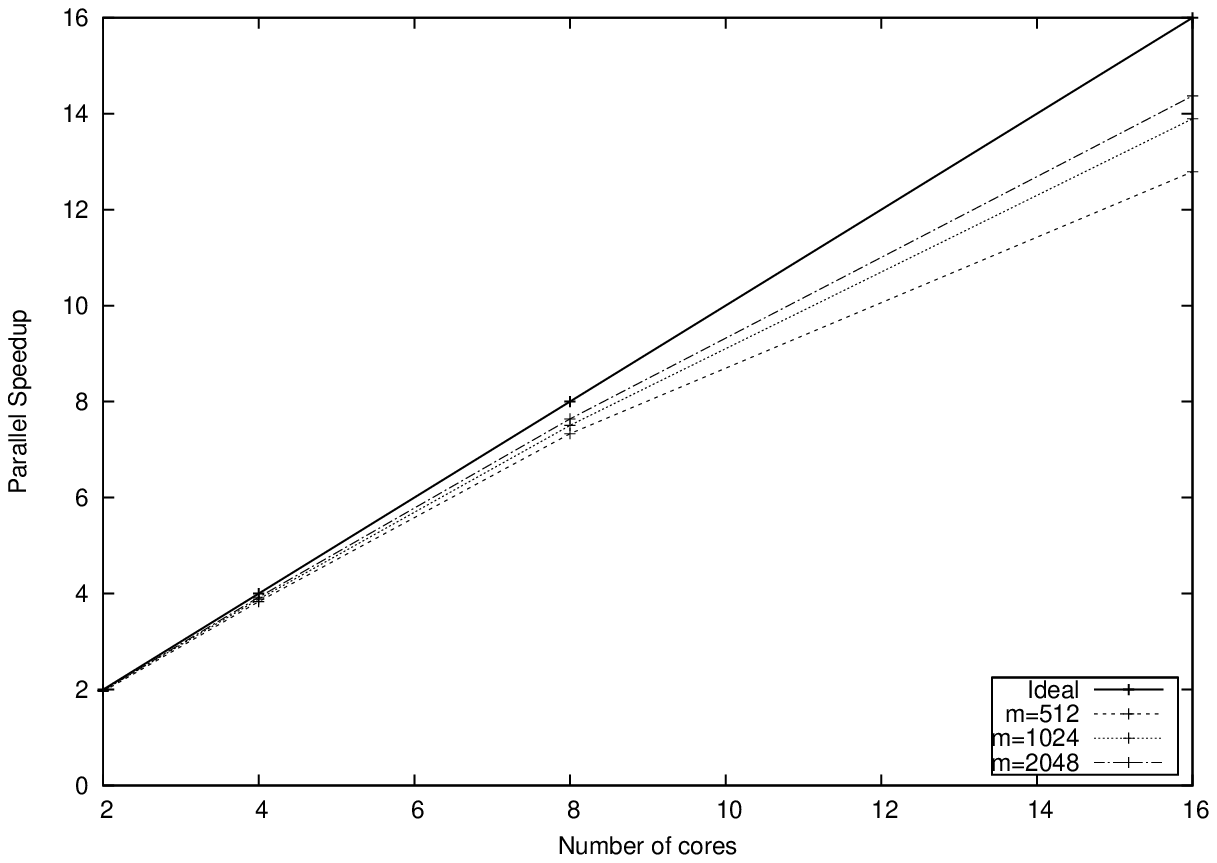}
	}
	\caption{Speedup results of the HHPCJaya algorithm for the five benchmark functions}
	\label{sp}
\end{figure}

In the second set of experiments, the parallel computation times and speedup of the HHCPJaya algorithm are analysed using the computational results in Tables~\ref{tab1} - ~\ref{tab4}. The time spent in the computation/optimisation process is measured in seconds. The speedup results of the HHCPJaya algorithm for five test functions and several values of $m$ and $n=64$ are shown against the number of cores in Figure~\ref{sp}. Similar speedup results are obtained for large population size $n=1024$ but these omitted owing to the space limitations.

From the results, it can observed that the execution time of the HHCPJaya algorithm decreases as the number of cores increases for all benchmark functions and any values of $conf\_h$ and $conf\_v$. Based on the speedup results, it is seen that there is a significant speedup gain compared to its sequential counterpart. Furthermore, the speedup is progressively improved by increasing the number of variables. Therefore, the performance results support the assertion the speedup of the proposed HHCP Jaya algorithm is approximately linear for higher dimensions and populations, and the scalability of the algorithm is approximately constant for all considered numbers of cores.

On the other hand, the performance results in Tables~\ref{tab1} -~\ref{tab4} indicate that the execution time of the HHCPJaya algorithm increases slightly when the values of the parameters $conf\_h$ and $conf\_v$ increase for each benchmark function. Similarly, there is a significant effect on the speedup results. For example, the speedup of the function $f_{2}$ for $m=1024$ decreases from 14.30 for $conf\_h=1$ and $conf\_v=1$ to 8.45 for $conf\_h=1$ and $conf\_v=32$ on 16 cores. Furthermore, the speedup also further decreases to 5.30 for $conf\_ h=2$ and $conf\_v=64$. This performance overhead is a result of the hyper-population approach, where each core has more subpopulations of small size to process. However, this additional time overhead is negligible for all test functions that may require greater number of iterations and significantly higher running times in the sequential algorithm to reach the global optimal solution. Furthermore, the running time of the HHCPJaya algorithm increases significantly when the population size and the values of the parameters $conf\_h$ and $conf\_v$  increase for each test function. This is due to the fact that each core processes subpopulations of large size and it is possible in increasing the cache misses.

As such, it is shown that the performance of the proposed multi-core HHCPJaya implementation with appropriate values of the parameters population size $n$, $conf\_h$ and $conf\_v$ is sufficiently efficient for large-scale global optimisation problems with time consuming function computations and large dimensions.

\subsection{Results for the Convergence Analysis}
In the final set of experiments, we present several results relating to the convergence ability of the proposed algorithm against the standard Jaya algorithm. The average number of iterations, average fitness value achieved, and time (in seconds) spent in the search process are presented in Table~\ref{iter1} for $m=512$ and Table~\ref{iter2} for $m=1024$. We report the fitness values achieved either in reaching the optimal solution with a fitness value of less than $1e^{-7}$ or after at most 10000 iterations. From these experimental results, it can observed that for all test functions and any values of the parameters $con\_h$ and $con\_v$, the average number of iterations of the proposed algorithm decreased as the number of cores or subpopulations increases  against the standard Jaya algorithm. Furthermore, we can see that the HHCPJaya algorithm achieves a better efficiency when the values of parameters  $con\_h$ and $con\_v$ increased for each test function, requiring a significantly lower number of iterations and search time to reach the global optimal solution. 
However, the search time of the standard Jaya algorithm is lower than the HHCPJaya algorithm on a small number of cores (i.e. up to 2 cores). This is because the proposed algorithm employes two additional procedures that are not present in the standard Jaya algorithm such as the two-level hierarchical cooperative search scheme and the updating phase on the subpopulations based on global solutions. By profiling of the code of the standard and parallel Jaya algorithms, we know that the updating phase on the population is the most computational intensive part of the code. In the standard Jaya algorithm, there is a call to the updating phase according to Algorithm~\ref{alg-jaya} (line 6), whereas in the proposed parallel algorithm, there are two calls to the updating phase according to Algorithm~\ref{algo7} (lines 8 and 16). The line 8 is the parallel updating phase on the subpopulations based on local solutions as is in original Jaya algorithm and the line 16 is an additional parallel updating phase on the subpopulations based on global solutions as we proposed in section~\ref{parjaya}. This has resulted in the parallel algorithm on one core takes at least double of the computation time of the standard Jaya algorithm, whereas the computation time of the parallel algorithm on two cores reduced by at least half. We also observe that there is a clear increase in the mean number of iterations and the search time of the HHCPJaya algorithm as the problem dimensions increases.

\begin{sidewaystable}
	\begin{tabular}{l|lll|lll|lll}
		\hline
		Function  &  \multicolumn{3}{c|}{$f_{1}$} &  \multicolumn{3}{c|}{$f_{2}$} &  \multicolumn{3}{c}{$f_{3}$}  \\ \hline
		Conficients & \multicolumn{3}{c|}{$conf\_h=1, conf\_v=1$} & \multicolumn{3}{c|}{$conf\_h=1, conf\_v=1$} & \multicolumn{3}{c}{$conf\_h=1, conf\_v=1$} \\ \hline
		Cores & Iterations & Mean f & Time (s) & Iterations & Mean f & Time (s) & Iterations & Mean f & Time (s) \\ \hline
		 1 (stad. Jaya) & 10000 & 2.14E+10 & 7.2316 & 10000 &  2.62E+09 & 11.8997 & 10000 & 1.69E+07 & 11.5505 \\
		1 &	10000 &	7.46E+03 &	17.2767 &	10000 &	1.31E+03 &	26.7580 &	10000 &	1.68E+01 &	26.2869\\
		2 &	8420 &	0 &	7.4166 &	10000 &	8.56E+02 &	13.6120 &	10000 &	5.57E+00 &	13.4156\\
		4 &	2135 &	0 &	0.9695 &	10000 &	8.63E+02 &	7.0042 &	3786 &	1.03E-01 &	2.6273\\
		8 &	1099 &	0 &	0.2696 &	10000 &	9.54E+02 &	3.7099 &	3945 &	3.83E+00 &	1.4767\\
		16 &	585 &	0 &	0.0889 &	10000 &	1.64E+03 &	2.1084 &	6180 &	6.96E+00 &	1.3504\\	
		\hline
		& \multicolumn{3}{c|}{$conf\_h=1, conf\_v=2$} & \multicolumn{3}{c|}{$conf\_h=1, conf\_v=16$} & \multicolumn{3}{c}{$conf\_h=1, conf\_v=4$} \\ \hline
		1 &	10000 &	1.60E-05 &	17.3865 &	10000 &	3.02E+03 &	28.5186 &	7017 &	7.35E-01 &	18,9191\\
		2 &	3002 &	0	& 2.6696	& 10000 &	1.75E+03 &	15.1859 &	3338 &	2.54E-01 &	4.6989\\
		4 &	1340 &	0	& 0.6221 &	10000 &	3.74E+02 &	8.1959 &	3615 &	0.54421 &	2.7536\\
		8 &	635 & 	0	& 0.1629 &	5418 &	0	 & 2.7085 &	860 &	0 &	0.3760\\
		16 &	314 &	0 &	0.0541 &	420	 &0 &	0.1608 &	382 &	0 &	0.1095\\		
		\hline
		& \multicolumn{3}{c|}{$conf\_h=2, conf\_v=4$} & \multicolumn{3}{c|}{$conf\_h=2, conf\_v=32$} & \multicolumn{3}{c}{$conf\_h=2, conf\_v=8$} \\ \hline
		1 &	2993 &	0	& 5.2702 &	10000 &	1.63E+03 &	30.5081 &	3589 &	2.95E-01 &	10.0094\\
		2 &	1340 &	0	& 1.2239 &	10000 &	2.55E+02 &	16.3263 &	2908 &	3.01E-01 &	4.3791\\
		4 &	633 &	0 &	0.3120 &	4855 &	0	 & 4.7367 &	913 &	0 &	0.7756\\
		8 &	312 &	0	& 0.0907 &	224 &	0 &	0.1626 &	371 &	0 &	0.1937\\
		16 &	206 &	0 &	0.0433 &	134 &	0 &	0.0914 &	248 &	0	 & 0.0968\\		
		\hline
	\end{tabular}
	\caption{Iteration number and function fitness values of the solutions obtained by the HHCPJaya algorithm for the benchmark functions ($n=64, m=512$)}
	\label{iter1}
\end{sidewaystable}

\addtocounter{table}{-1}
\begin{sidewaystable}
	\begin{tabular}{l|lll|lll}
		\hline
		Function  &  \multicolumn{3}{c|}{$f_{4}$} &  \multicolumn{3}{c}{$f_{5}$}  \\ \hline
		Conficients & \multicolumn{3}{c|}{$conf\_h=1, conf\_v=1$} & \multicolumn{3}{c}{$conf\_h=1, conf\_v=1$}  \\ \hline
		Cores & Iterations & Mean f & Time (s)  & Iterations & Mean f & Time (s)  \\ \hline
		 1 (stad. Jaya) & 10000 & 2.01E+08 & 14.8218 & 10000 &  6.55E+14 & 7.7129 \\
		1	& 10000 &	6.81E+01 &	32.5719 &	10000 &	4.41E+07 &	19.1551\\
		2	& 7090 &	0 &	11.7298 &	10000 &	5.08E+06 &	9.7537\\
		4	& 1798 &	0 &	1.5294 &	10000 &	1.13E+01 &	5.0136\\
		8	& 928 &	0	 & 0.4198 &	10000 &	2.35E+06 &	2.6835\\
		16 &	491 &	0 &	0.1304 &	10000 &	3.72E+09 &	1.5874\\	
		\hline
		& \multicolumn{3}{c|}{$conf\_h=1, conf\_v=2$} & \multicolumn{3}{c}{$conf\_h= 1, conf\_v=16$}  \\ \hline
		1	& 9962 &	1.00E-06 &	32.6004 &	10000 &	1.82E+07 &	20.7428\\
		2	& 2523 &	0	& 4.2187 &	10000 &	4.99E+00 &	11.3452\\
		4 &	1128 &	0 &	0.9785 &	9988 &	3.37E-04 &	5.9832\\
		8 &	535 &	0 &	0.2527 &	6937 &	0	 &2.6014\\
		16 &	265 &	0 &	0.0767 &	929 &	0 &	0.26007\\
		\hline
		& \multicolumn{3}{c|}{$conf\_h=2, conf\_v=4$} & \multicolumn{3}{c}{$conf\_h=2,conf\_v=32$}  \\ \hline
		1 &	2524 &	0	& 8.3505	& 10000	& 2.14E-02 &	22.5212\\
		2	& 1129 &	0	& 1.9304 &	9385 &	1.00E-06 &	11.1102\\
		4 &	533 &	0	& 0.4838 &	4971 &	0	 & 3.6435\\
		8 &	262 &	0 &	0.1344 &	754 &	0 &	0.3948\\
		16&	173 &	0	 & 0.0587 &	* &	* &	*\\
		\hline
	\end{tabular}
	\caption{(continued)}
\end{sidewaystable}

\begin{sidewaystable}
	\begin{tabular}{l|lll|lll|lll}
		\hline
		Function  &  \multicolumn{3}{c|}{$f_{1}$} &  \multicolumn{3}{c|}{$f_{2}$} &  \multicolumn{3}{c}{$f_{3}$}  \\ \hline
		Conficients & \multicolumn{3}{c|}{$conf\_h=1, conf\_v=1$} & \multicolumn{3}{c|}{$conf\_h=1, conf\_v=1$} & \multicolumn{3}{c}{$conf\_h=1, conf\_v=1$} \\ \hline
		Cores & Iterations & Mean f & Time (s) & Iterations & Mean f & Time (s) & Iterations & Mean f & Time (s) \\ \hline
		1 (stad. Jaya) & 10000 &  1.50E+11 & 14.4011 & 10000 &  8.62E+09 & 23.7351 & 10000 & 1.89E+07 & 23.0146 \\
		1 &	10000	& 9.61E+04 &	34.4821 &	10000 &	4.91E+03 &	53.4517 &	10000 &	1.90E+01 &	52.4137\\
		2 &	10000 &	4.56E+02 &	17.4031 &	10000 &	2.04E+03 &	26.9758	& 10000 &	1.22E+01 &	26.5092\\
		4 &	4641 &	0 &	4.1245 &	10000 &	1.73E+03 &	13.7954 &	10000 &	2.67E+00	& 13.5782\\
		8	& 2064 &	0	 & 0.9645 &	10000 &	2.12E+03 &	7.2151 &	3519 &	1.62E+00 &	2.5205\\
		16 &	1228 &	0	& 0.3191 &	10000 &	5.51E+03 &	3.9266 &	8326 &	1.38E+01 &	3.2905\\	
		\hline
		& \multicolumn{3}{c|}{$conf\_h=1, conf\_v=2$} & \multicolumn{3}{c|}{$conf\_h=1, conf\_v=32$} & \multicolumn{3}{c}{$conf\_h=1, conf\_v=8$} \\ \hline
		1 &	10000 &	1.56E+03 &	34.5152 &	10000 &	6.13E+03 &	57.0899	& 6822 &	1.09E+00 &	36.7400\\
		2 &	6359 &	0	& 11.1289 &	10000 &	3.54E+03 &	30.3478 &	5197 &	1.05E+00 &	14.5919\\
		4 &	2271 &	0	& 2.0473 &	10000 &	7.76E+02 &	16.4527 &	4762 &	1.00E+00 &	7.2409\\
		8 &	1237 &	0	& 0.5909 &	5989 &	0.51538 &	5.9503 &	1068 &	0	 & 0.9245\\
		16 &	628 &	0 &	0.1743 &	458 &	0 &	0.3358 &	435 &	0 &	0.2381\\		
		\hline
		& \multicolumn{3}{c|}{$conf\_h=2, conf\_v=4$} & \multicolumn{3}{c|}{$conf\_h=2, conf\_v=64$} & \multicolumn{3}{c}{$conf\_h=2, conf\_v=16$} \\ \hline
		1 &	6375 &	0	& 22.1913 &	10000 &	3.31E+03 &	60.3680	 &5492 &	7.84E-01 &	30.6376\\
		2	& 2269 &	0 &	4.0494 &	10000 &	5.25E+02 &	32.5721 &	5174 &	8.32E-01 &	15.5810\\
		4 &	1230 &	0	 & 1.1497 &	5845	& 4.97E-02 &	11.3880 &	1063 &	0 &	1.8080\\
		8	& 624 &	0 &	0.3247 &	265 &	0 &	0.3798 &	439 &	0 &	0.4572\\
		16 &	392 &	0	& 0.1249 &	143 &	0 &	0.1850 &	273 &	0	 & 0.2029\\	
		\hline
	\end{tabular}
	\caption{Iteration number and function fitness values of the solutions obtained by the HHCPJaya algorithm for the benchmark functions ($n=64, m=1024$)}
	\label{iter2}
\end{sidewaystable}

\addtocounter{table}{-1}
\begin{sidewaystable}
	\begin{tabular}{l|lll|lll}
		\hline
		Function  &  \multicolumn{3}{c|}{$f_{4}$} &  \multicolumn{3}{c}{$f_{5}$}  \\ \hline
		Conficients & \multicolumn{3}{c|}{$conf\_h=1, conf\_v=1$} & \multicolumn{3}{c}{$conf\_h=1, conf\_v=1$}  \\ \hline
		Cores & Iterations & Mean f & Time (s) & Iterations & Mean f & Time (s)\\ \hline
		 1 (stad. Jaya) & 10000 & 1.35E+09 & 29.6195 & 10000 &  1.02E+16 &  15.3950 \\
		1 &	10000 &	8.69E+02 &	65.0487 &	10000 &	3.90E+09 &	38.2537\\
		2 &	10000 &	4.82E+00 &	32.8374 &	10000 &	2.83E+09 &	19.3023\\
		4 & 3906 &	0 &	6.5417 &	10000 &	2.26E+02 &	9.8480\\
		8 &	1742 &	0 &	1.5296 &	10000 &	9.03E+01 &	5.1767\\
		16 &	1037 &	0 &	0.4978 &	10000 &	7.50E+09 &	2.8404\\		
		\hline
		& \multicolumn{3}{c|}{$conf\_h=1, conf\_v=2$} & \multicolumn{3}{c}{$conf\_h=1, conf\_v=32$}  \\ \hline
		1 &	10000 &	1.68E+01 &	65.1190 &	10000 &	7.07E+03 &	41.5140\\
		2 &	5332 &	0	 & 17.5800 &	10000 &	8.21E+00 &	22.6726\\
		4 &	1920 &	0 &	3.2498 &	10000 &	2.97E-02 &	12.0029\\
		8	& 1045 &	0	 & 0.9351 &	7385 &	0	& 5.5559\\
		16 &	531 &	0& 	0.2679 &	1025 &	0 &	0.5537\\	
		\hline
		& \multicolumn{3}{c|}{$conf\_h=2, conf\_v=4$} & \multicolumn{3}{c}{$conf\_h=2, conf\_v=64$}  \\ \hline
		1 &	5318 &	0	& 34.8347 &	10000 &	5.04E-01 &	45.1381 \\
		2	& 1911 &	0	& 6.3922 &	9979 &	4.80E-05 &	23.7781\\
		4	& 1040 &	0	& 1.8123 &	5329 &	0	& 7.8162\\
		8	& 528 &	0 &	0.4975 &	870 &	0	 &0.9055\\
		16 &	330 &	0 &	0.1826 &	*	& *	 & *\\
		\hline
	\end{tabular}
	\caption{(continued)}
\end{sidewaystable}

Based on these results, we can conclude that the convergence rate is high and execution time is fast for the HHCPJaya algorithm when the maximum number of small subpopulations is employed. In other words, the HHCPJaya algorithm is efficient for a large number of small subpopulations, because it gives the optimal solution with a smaller number of iterations and running time. For example, the HHCPJaya algorithm for the Rastrigin function and $m=512$ produces the optimal solution with fewest iterations (i.e., 134) and the lowest execution time (i.e., 0.0914 seconds) on 16 cores when the values of the parameters $con\_h$ and $con\_v$  are 2 and 32, respectively (or the number and the sizes of the subpopulations are 16384 and $2 \times 1$, respectively). Note that the running time of the proposed parallel implementation using the hyper-population approach is much lower compared to the standard Jaya algorithm, which requires more time to reach the global optima. Another reason for the fast convergence rate and execution time of the HHCPJaya algorithm is the two-level hierarchical cooperative search scheme among the subpopulations.
When the size of the subpopulations obtained by the hyper-population approach is sufficiently small, the locality of computations increases, minimising the cache misses. Therefore, the computations on these small subpopulations are cache-friendly, helping to improve the performance. This works well across complex memory hierarchies of multicore platforms.

Finally, it is clear that there is an inverse relationship between the number of iterations and the values of $con\_h$ and $con\_v$ for the algorithm to produce the best results with the lowest search times. Therefore, the HHCPJaya algorithm for a given function and number of cores can produce an optimal solution efficiently by increasing the number of subpopulations and decreasing the number of iterations or function evaluations.

\section{Conclusions}
\label{conclusions}
We proposed a new HHCPJaya algorithm for solving large-scale global optimisation problems, which incorporates three characteristics: the hyper-population approach, a two-level local and global hierarchical cooperative search scheme among the subpopulations, and an updating phase for solutions on the respective local subpopulations based on the global solutions. Moreover, we implemented the proposed algorithm on a multi-core platform using the OpenMP programming interface, and we also evaluated its performance in terms of solution quality, running time, and convergence ability on a set of benchmark functions provided for the CEC2013 Special Session on a large-scale global optimisation problem. The computational experiments showed that the HHCPJaya algorithm significantly improves on the performance of the original Jaya algorithm, and is also a fast and fully parallelisable algorithm. The results indicated that the parameters of the hyper-population approach (i.e., the number and the sizes of subpopulations) play a major role in the quality of solutions, convergence rate, and number of iterations. Finally, it was shown that the HHCPJaya algorithm with a large number of small subpopulations and minimum number of function evaluations (or number of iterations) is highly efficient for large-scale global optimisation problems. This is because an optimal solution is produced with a fast convergence rate and much lower running time compared with the sequential Jaya algorithm, which requires more iterations and longer running times to reach global optima. Therefore, the proposed hierarchical hyper-population cooperative parallel Jaya algorithm provides high quality solutions, and improves the convergence rate with a smaller computational effort.

Finally, a number of directions for future research can be identified. First, the performance comparison of the proposed HHCPJaya algorithm could be extended to more complex functions from the CEC 2013 test suite, constrained benchmark optimisation problems. Second, we are interested in applying of the proposed parallel algorithm to some real-life applications involve optimisation of a large number of variables. For example, in multidimensional scaling for analysis and visualization of multidimensional data might reduced to a difficult and high dimensional global optimisation problem \cite{ZILINSKAS2006211}. Furthermore, in shape optimisation a relatively large number of shape design variables is often used to represent complex shapes \cite{Olhofer2002, SKINNER2017}. Finally, the HHCPJaya algorithm the hierarchical hyper-population cooperative approach could be efficiently implemented on other parallel programming paradigms, such as MPI, CUDA, and hybrid schemes such as MPI-OpenMP, MPI-CUDA, and MPI-OpenMP-CUDA.

\section*{Acknowledgements}
The author would like to thank the anonymous reviewers for helpful comments, which have improved the presentation of this paper.
The computational experiments reported in this paper were performed at the Parallel and Distributed Processing Laboratory (PDP Lab) of the Department of Applied Informatics, University of Macedonia. The author would also like to thank the personnel of the PDP Lab.

\bibliographystyle{gPAA}
\bibliography{gPAAguide}

\end{document}